\documentclass[twocolumn]{aastex631}

\usepackage{array,multirow}

\shorttitle{The Spacing between $^{13}$CO structures within $^{12}$CO molecular clouds}
\shortauthors{Yuan et al.}
\graphicspath{{./}{figures/}}

\begin{document}
\title{On the Spatial Distribution of $^{13}$CO Structures within $^{12}$CO Molecular Clouds}

\correspondingauthor{Ji Yang}
\email{jiyang@pmo.ac.cn}
\author[0000-0003-0804-9055]{Lixia Yuan}
\affiliation{Purple Mountain Observatory and Key Laboratory of Radio Astronomy, Chinese Academy of Sciences, \\
10 Yuanhua Road, Qixia District, Nanjing 210033, PR China}

\author[0000-0001-7768-7320]{Ji Yang}
\affiliation{Purple Mountain Observatory and Key Laboratory of Radio Astronomy, Chinese Academy of Sciences, \\
10 Yuanhua Road, Qixia District, Nanjing 210033, PR China}

\author[0000-0002-7489-0179]{Fujun Du}
\affiliation{Purple Mountain Observatory and Key Laboratory of Radio Astronomy, Chinese Academy of Sciences, \\
10 Yuanhua Road, Qixia District, Nanjing 210033, PR China}

\author[0000-0001-8315-4248]{Xunchuan Liu} 
\affiliation{Shanghai Astronomical Observatory, Chinese Academy of Sciences, PR China}

\author[0000-0002-0197-470X]{Yang Su}
\affiliation{Purple Mountain Observatory and Key Laboratory of Radio Astronomy, Chinese Academy of Sciences, \\
10 Yuanhua Road, Qixia District, Nanjing 210033, PR China}

\author[0000-0003-4586-7751]{Qing-Zeng Yan}
\affiliation{Purple Mountain Observatory and Key Laboratory of Radio Astronomy, Chinese Academy of Sciences, \\
10 Yuanhua Road, Qixia District, Nanjing 210033, PR China}

\author{Xuepeng Chen}
\affiliation{Purple Mountain Observatory and Key Laboratory of Radio Astronomy, Chinese Academy of Sciences, \\
10 Yuanhua Road, Qixia District, Nanjing 210033, PR China}

\author[0000-0002-3904-1622]{Yan Sun}
\affiliation{Purple Mountain Observatory and Key Laboratory of Radio Astronomy, Chinese Academy of Sciences, \\
10 Yuanhua Road, Qixia District, Nanjing 210033, PR China}

\author[0000-0003-2549-7247]{Shaobo Zhang}
\affiliation{Purple Mountain Observatory and Key Laboratory of Radio Astronomy, Chinese Academy of Sciences, \\
10 Yuanhua Road, Qixia District, Nanjing 210033, PR China}

\author[0000-0003-2418-3350]{Xin Zhou}
\affiliation{Purple Mountain Observatory and Key Laboratory of Radio Astronomy, Chinese Academy of Sciences, \\
10 Yuanhua Road, Qixia District, Nanjing 210033, PR China}

\author[0000-0002-8051-5228]{Yuehui Ma}
\affiliation{Purple Mountain Observatory and Key Laboratory of Radio Astronomy, Chinese Academy of Sciences, \\
10 Yuanhua Road, Qixia District, Nanjing 210033, PR China}

\begin{abstract}
We look into the 2851 $^{12}$CO molecular clouds harboring $^{13}$CO structures to reveal 
the distribution of the projected angular separations and radial velocity separations 
between their internal $^{13}$CO structures. 
The projected angular separations are determined using the minimal spanning tree algorithm. 
We find that $\sim$ 50$\%$ of the angular separations fall in a narrow range of $\sim$ 3 -- 7 arcmin with a median of $\sim$ 5 arcmin, 
and the corresponding radial velocity separations mainly range from $\sim$ 0.3 km s$^{-1}$ to 2.5 km s$^{-1}$.
The mean and standard deviation of the angular separations of the internal $^{13}$CO structures within $^{12}$CO clouds 
appear to be universal, independent of the $^{12}$CO cloud angular areas and the counts of their internal $^{13}$CO structures.
We also reveal a scaling relation between the $^{12}$CO cloud angular area and its harbored $^{13}$CO structure count.
These results suggest there is a preferred angular separation between $^{13}$CO structures in these $^{12}$CO clouds, 
considering the distance effects. 
According to that, we propose an alternative picture for the assembly and destruction of molecular clouds: there is a fundamental 
separation for the internal structures of molecular clouds, the build-up and destruction of molecular clouds proceeds under 
this fundamental unit.  
\end{abstract}


\section{Introduction} \label{sec:intro}

Molecular clouds (MCs) are the densest (n $>$ 30 cm$^{-3}$), darkest, and coldest ($\sim$ 10 K -- 20 K) 
molecular phase of the interstellar medium (ISM). 
They are linked with star formation, planet formation, and galaxy evolution. 
Hence it is essential to understand the nature of MCs, passing through an understanding on 
their hierarchical structures, physical properties, and dynamics. 
Several typical mechanisms on the formation and evolution of MCs have been proposed, 
including large-scale gravitational instabilities \citep{Lin1964, Goldreich1965}, 
turbulent flows \citep{Vazquez1995, Passot1995, Ballesteros1999, Vazquez2006, Heitsch2006, Beuther2020}, 
and agglomeration of smaller clouds \citep{Oort1954, Field1965, Dobbs2014}, 
however many questions are still open. For instance, giant molecular clouds usually 
exhibit complex and hierarchical structures,  
inside which smaller and denser structures or clumps are found at different levels of the hierarchy, 
it is still not clear that how these internal substructures and dense gas gather in MCs.


Filamentary structures are widespread in molecular clouds \citep{Andre2010, Molinari2010}. 
The filament populations have a wide range of properties in terms of lengths, 
aspect ratios, widths, and masses \citep{Arzoumanian2011, Li2013, Hacar2013, Wang2016, 
Mattern2018, Hacar2018, Hacar2022}, they are also known to play an important 
role in star formation, considering the observed tendency of pre-stellar 
cores and young stellar objects (YSOs) to be associated with filaments \citep{Myers2009, Polychroni2013, Schisano2014, Andre2014, 
Andre2016, Konyves2015, Hacar2018, Konyves2020, Yuan2020}.  
Our previous works in \cite{Yuan2021} (Paper I) classified 18,190 MCs, which were identified by $^{12}$CO lines data 
from the Milky Way Imaging Scroll Painting (MWISP) survey \citep{Su2019}, into filaments and nonfilaments \citep{Yuan2021}. 
In that work, it found that filaments making up $\sim$ 10$\%$ of the total number, 
contribute $\sim$ 90$\%$ of the total integrated fluxes of $^{12}$CO line emission.
In addition, \cite{Yuan2022} (Paper II) extracted $^{13}$CO gas structures within each of 18,190 $^{12}$CO clouds using 
the $^{13}$CO lines from the MWISP survey. 
Among that, there are 2851 $^{12}$CO clouds ($\sim$ 15$\%$) having $^{13}$CO gas structures. 
Combining the results from Paper I and II, 
we find that the MCs classified as filaments tend to have larger spatial areas and more substructures traced by $^{13}$CO lines. 
However, what kind of a relation does exist between the spatial scales of $^{12}$CO clouds and 
their internal $^{13}$CO structures? 
Answering this question may help to understand how molecular gas gathers to form stars on its way from MCs to substructures.  

Among the 2851 $^{12}$CO MCs having $^{13}$CO structures, 
$\sim$ 60$\%$ of these $^{12}$CO clouds have a single $^{13}$CO structure, 
$\sim$ 15$\%$ of them have double $^{13}$CO structures, 
and the rest $\sim$ 20$\%$ of them have multiple (more than two) $^{13}$CO structures \citep{Yuan2022}.  
These results indicate the distribution of high-density gas in MCs is inhomogenous. 
Using $^{13}$CO line emission within $^{12}$CO clouds, 
we have provided the high-density gas content in these $^{12}$CO clouds \citep{Yuan2022}. 
However, the spatial distributions of the $^{13}$CO structures within $^{12}$CO clouds are 
still not well investigated. This information may provide essential clues to understand 
how the dense gas content and the internal sub-structures develop in the MCs, to reveal 
the building-up process of MCs.   

In this paper, we aim to reveal the spatial distribution and the relative motion between 
the internal $^{13}$CO structures within 2851 $^{12}$CO clouds.  
Section 2 mainly describes the $^{12}$CO and $^{13}$CO lines data, the $^{12}$CO molecular cloud samples, and 
their internal $^{13}$CO structures. 
Section 3 presents the results, including the physical properties of these individual $^{13}$CO structures, 
the distributions of the projected angular separations and radial velocity separations between $^{13}$CO 
structures within these $^{12}$CO clouds. 
Section 4 mainly discusses the observational effects on the observed $^{13}$CO structure separations and 
also reveal the scaling relations between the internal $^{13}$CO structures and their natal $^{12}$CO clouds. 
We conclude in Section 5 with our results. 

\section{Data}
\subsection{$^{12}$CO and $^{13}$CO lines data from MWISP survey}
The MWISP survey is an ongoing northern Galactic plane CO survey, 
which utilizes the 13.7m telescope at Delingha, China, 
to observe the J= 1-0 transitions of the $^{12}$CO, $^{13}$CO, and C$^{18}$O lines, simultaneously. 
A detailed description of the survey is given in \cite{Su2019}. 
The typical noise temperature is $\sim$ 250 K for $^{12}$CO line and $\sim$ 140 K for $^{13}$CO/C$^{18}$O lines. 
The half-power beamwidth (HPBW) is about 50$^{\prime \prime}$ at 115 GHz.
The velocity resolution of $^{13}$CO lines is about 0.17 km s$^{-1}$. 
An rms noise level of $\sim$ 0.5 K for $^{12}$CO lines and $\sim$ 0.3 K for the $^{13}$CO lines 
are achieved.

In this work, we focus on the $^{12}$CO and $^{13}$CO line emission in the second Galactic quadrant with 
$104^{\circ}.75 < l < 150^{\circ}.25$, $|b| < 5^{\circ}.25$, and $-$95 km s$^{-1}$  $<$ V$_{\rm LSR}$ $<$ 25 km s$^{-1}$. 
The $^{12}$CO and $^{13}$CO lines emission data in this region have been published in paper I and II, respectively. 

\subsection{$^{12}$CO molecular clouds and $^{13}$CO structures}

A catalog of 18,190 $^{12}$CO molecular clouds has been identified from the $^{12}$CO line emission in the above region, 
using the Density-based Spatial Clustering of Applications with Noise (DBSCAN) algorithm \citep{Ester1996, Yan2021}. 
The DBSCAN algorithm extracts a set of contiguous voxels in the 
position-position-velocity (PPV) cube with $^{12}$CO line intensities above a certain threshold as a molecular cloud. 
The line intensity threshold is determined by the parameter of `cutoff'. 
The connectivity of extracted structures is confined by two parameters of $\epsilon$ and MinPts. 
A core point within the extracted consecutive structure, 
whose adjacent points within a certain radius have to exceed a number threshold. 
The `MinPts' determines the threshold of the number of adjacent points and $\epsilon$ 
determines the radius of the adjacence.  
The border points are the points inside the $\epsilon$-neighborhood of core points, 
but do not contain the `MinPts' neighbors in its $\epsilon$-neighborhood \citep{Ester1996}. 
The parameters of cutoff = 2$\sigma$ ($\sigma$ is the rms noise, 
whose value is $\sim$ 0.5 K for $^{12}$CO line emission), 
MinPts=4, and $\epsilon$=1 are adopted in the DBSCAN algorithm for $^{12}$CO clouds identification, 
as suggested in \cite{Yan2020}.  
In addition, the post-selection criteria are also utilized to avoid noise contamination. 
For each extracted cloud, it needs to satisfy:(1) the total voxel number is larger than 16; 
(2) the peak intensity is higher than the `cutoff' adding 3$\sigma$; 
(3) the angular area is larger than one beam size (2$\times$2 pixels $\sim$ 1 arcmin); 
and (4) the number of velocity channels is larger than 3. 
The observational effects, including the finite angular resolution and sensitivity of the observed spectral lines data, 
on the molecular cloud samples have been systematically investigated in \cite{Yan2022}.
In addition, the samples extracted by the DBSCAN algorithm also have been compared with 
those from other algorithms in \cite{Yan2022, Yuan2022}.

$^{13}$CO structures are defined as a set of contiguous voxels in the PPV space 
with $^{13}$CO line intensities above a certain threshold, which are extracted using the 
DBSCAN algorithm within $^{12}$CO cloud boundaries.  
The used DBSACN parameters are identical to the above parameters for $^{12}$CO cloud extraction, 
except for the post-selection criteria of the peak intensities higher than the `cutoff' adding 2$\sigma$, 
$\sigma$ is $\sim$ 0.25 K for $^{13}$CO line emission. 
Among the total 18,190 $^{12}$CO clouds, 2851 $^{12}$CO clouds are identified to 
have $^{13}$CO structures in our paper II. 
The extracted $^{12}$CO line data for 18,190 $^{12}$CO clouds and the extracted 
$^{13}$CO structures within the 2851 $^{12}$CO clouds are available at DOI:\href{https://doi.org/10.57760/sciencedb.j00001.00427}{10.57760/sciencedb.j00001.00427} 
\citep{Yuan2022b}. 
In this work, we focus on these 2851 $^{12}$CO molecular clouds and their internal $^{13}$CO structures to investigate 
the spatial distribution of these internal $^{13}$CO structures. 

\section{Results}

\subsection{Physical properties of $^{13}$CO structures}

There are a total of 9566 $^{13}$CO structures within the 2851 $^{12}$CO molecular clouds. 
Among that, 1848 $^{13}$CO structures (19.3$\%$) are within the $^{12}$CO clouds having a single $^{13}$CO structure, 
886 $^{13}$CO structures (9.3$\%$) are from $^{12}$CO clouds with double $^{13}$CO structures, 
and 6832 $^{13}$CO structures (71.4$\%$) are in the $^{12}$CO clouds having more than two $^{13}$CO structures (multiple). 
Based on that, the whole 9,566 $^{13}$CO structures are separated into three regimes, 
i.e. single, double, and multiple. 

\begin{figure*}[ht]
    \plotone{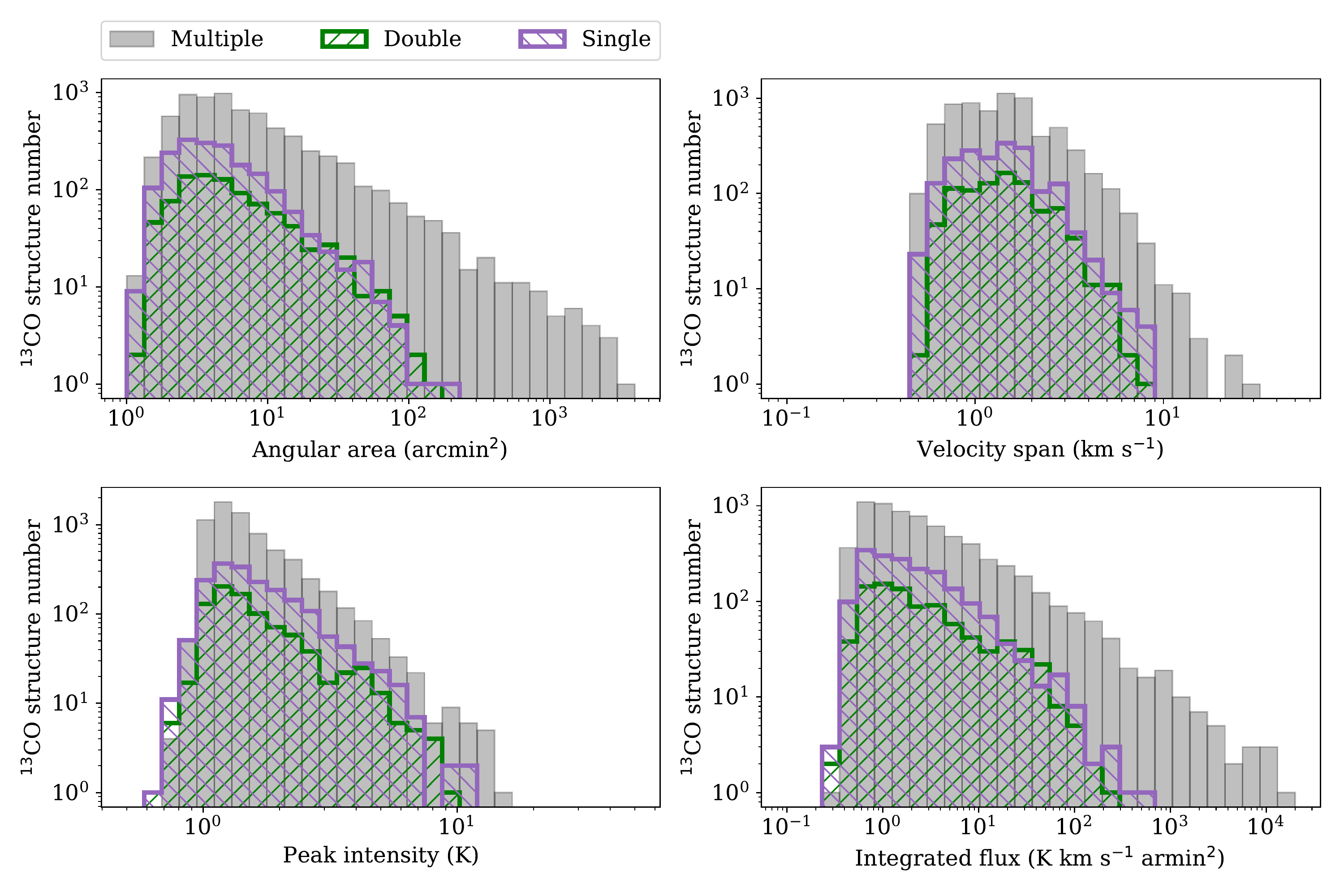} 
    \caption{The number distributions of angular area, velocity span, peak intensity, and integrated flux 
    of each individual $^{13}$CO structure. The purple, green, and gray histograms represent $^{13}$CO structures 
    that are within the $^{12}$CO clouds having single, double, and multiple (more than two) $^{13}$CO structures, respectively. \label{fig:fstr_dist}}
\end{figure*}

Figure \ref{fig:fstr_dist} presents the distributions of physical properties of these $^{13}$CO structures in three regimes. 
The properties include angular areas, velocity spans, peak intensities, and integrated fluxes of the $^{13}$CO line emission.   
Their number distributions for $^{13}$CO structures in three regimes are similar, 
except for that the `multiple' regime exists the $^{13}$CO structures having larger values in the angular areas ($\gtrsim$ 200 arcmin$^{2}$), 
velocity spans ($\gtrsim$ 10 km s$^{-1}$), and integrated fluxes ($\gtrsim$ 800 K km s$^{-1}$ arcmin$^{2}$). 
That suggests $^{13}$CO gas structures in the `multiple' system can 
develop into structures having larger spatial scales and higher masses. 

\begin{figure*}
    \plotone{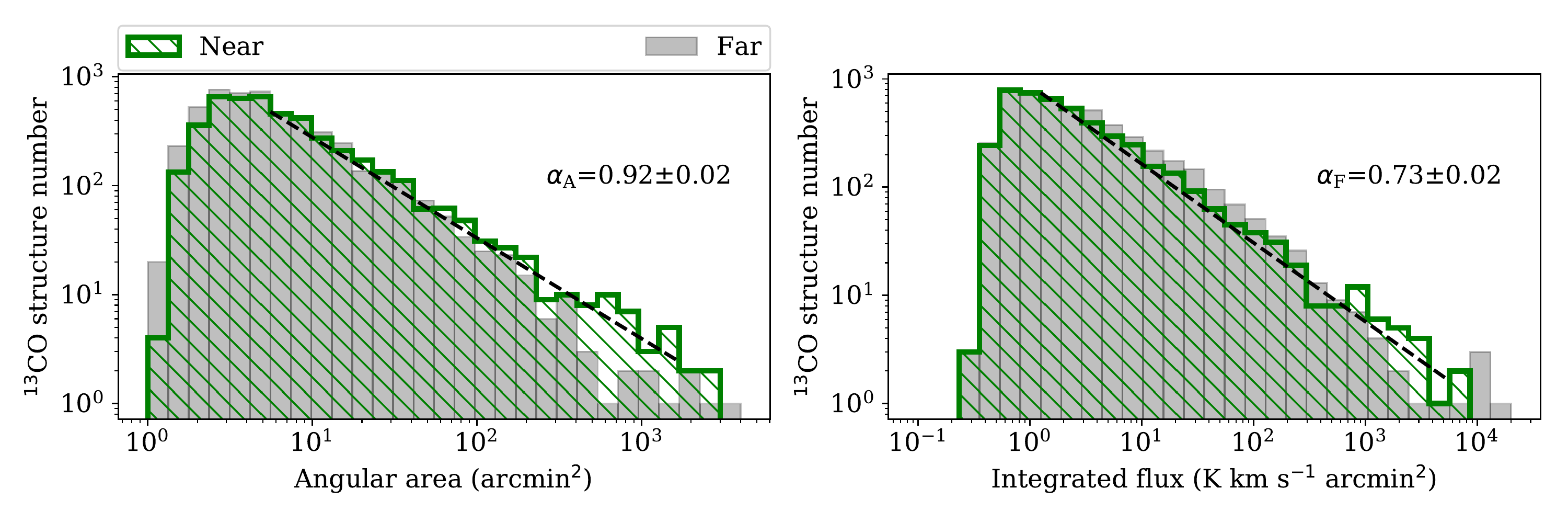}
    \caption{The number distributions of the angular areas (left) and integrated fluxes (right) for $^{13}$CO structures in 
    the near and far groups, respectively. The dashed lines fit the power-law distribution, 
    $dN/dX \propto X^{-(\alpha_{X}+1)}$, for the anguar areas and integrated fluxes of $^{13}$CO structures in the near group. 
    The fitted $\alpha_{X}$ is noted in each panel. \label{fig:fnf}}
\end{figure*}

Based on the spiral structure model of the Milky Way, 
$^{12}$CO clouds were roughly divided into two groups, i.e. near and far, by a V$_{\rm LSR}$ threshold of -30 km s$^{-1}$, 
in the paper I and II. 
The $^{12}$CO clouds in the near group have central velocities ranging from -30 km s$^{-1}$ to 25 km s$^{-1}$, 
and those in the far group are from -95 km s$^{-1}$ to -30 km s$^{-1}$.  
The $^{13}$CO structures are also divided into the near or far group, following their natal $^{12}$CO clouds. 
The kinematical distances of these $^{12}$CO molecular clouds, estimated using the Bayesian 
distance calculator in \cite{Reid2016}, mainly distribute on $\sim$ 0.5 kpc for clouds 
in the Local region and $\sim$ 2 kpc for those in the Perseus region.  
Considering these kinematical distances, the physical scale for the molecular cloud with an angular size of 1$^{\prime}$ in the Local region 
is $\sim$ 0.15 pc, and the value is $\sim$ 0.6 pc for that in the Perseus region. 

Figure \ref{fig:fnf} presents the number distributions of the angular areas and integrated fluxes 
of the $^{13}$CO structures, which are in the near and far groups, respectively. 
We find that the angular-area distribution for the observed $^{13}$CO structures in the near and far groups is close, 
as well as the distributions of their integrated fluxes.  
For the $^{13}$CO structures in near group, we fit their power-law distributions as $dN/dX \propto X^{-(\alpha_{X}+1)}$, 
the fitted $\alpha_{A}$ is 0.92 for the angular areas ($A$) and $\alpha_{F}$ is 0.73 for the integrated fluxes ($F$). 

In previous works, these localized and discrete sub-units in the observed intensity distribution 
were usually called clumps, such as the clumpy structures traced by the $^{13}$CO emission in \cite{Blitz1986, Williams1994, Williams1995, Kramer1998}. 
These clump masses, which were within various molecular clouds and 
were identified using different methods (either Guassclumps or Clumfind), 
agree in showing a clump mass spectrum following a power law form 
$dN/dM \propto M^{-\alpha}$ with a spectral index $\alpha$ between 1.2 and 1.9 \citep{Williams1994, Williams1995, Kramer1998}. 
Assuming the distances for $^{13}$CO structures in the same group are close, 
the mass of a $^{13}$CO structure will be proportional to its integrated flux, 
whose number distribution follows a power law with an exponent of 1.73$\pm$0.22. 
This value of 1.73$\pm$0.22 is in the range of 1.2 -- 1.9, 
thus the results of $^{13}$CO structures in Figure \ref{fig:fnf} are 
comparable with the previous clump mass spectrum derived in \cite{Williams1994, Williams1995, Kramer1998}.

\begin{figure*}[h]
    \plotone{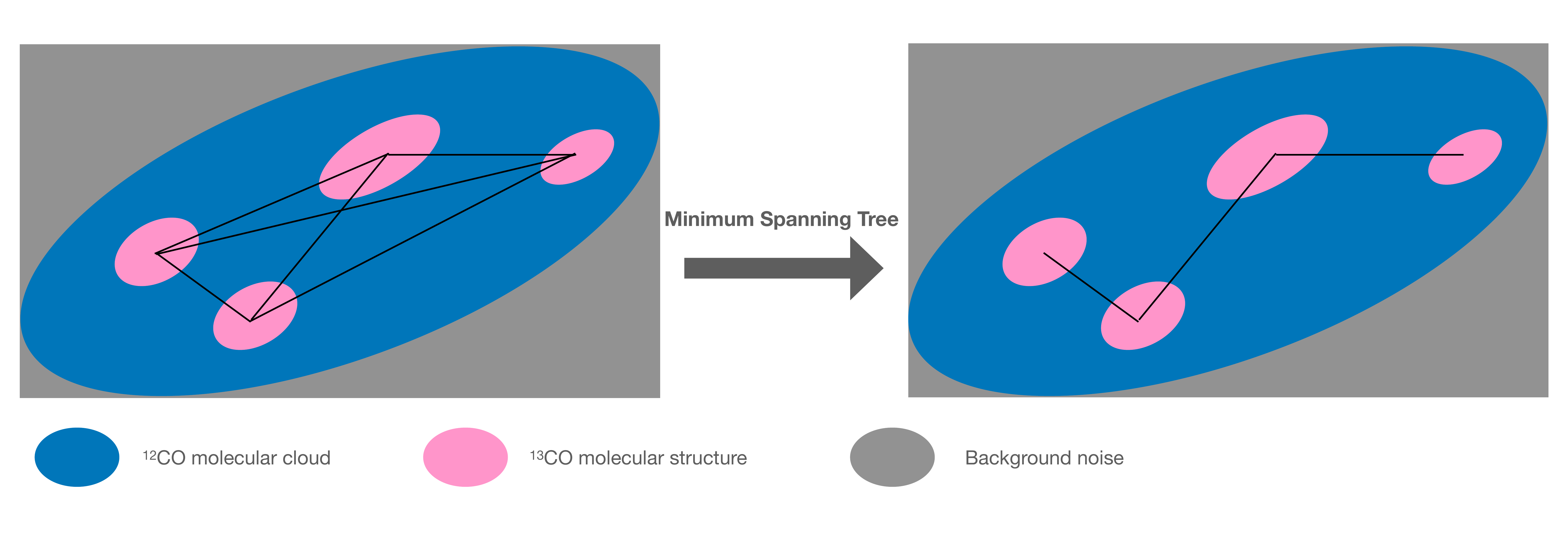} 
    \caption{A sketch demonstrating the computation of a minimum spanning tree (MST) over a $^{12}$CO cloud 
    having four $^{13}$CO structures. \textbf{Left panel}: The black lines show the connected graph 
    between each two of the four $^{13}$CO structures. \textbf{Right panel}: The black lines show an MST, 
    which connects the whole $^{13}$CO structures and minimizes the total sum of projected angular 
    separations between $^{13}$CO structures. \label{fig:fMST}}
\end{figure*}

\begin{figure*}[h]
    \plotone{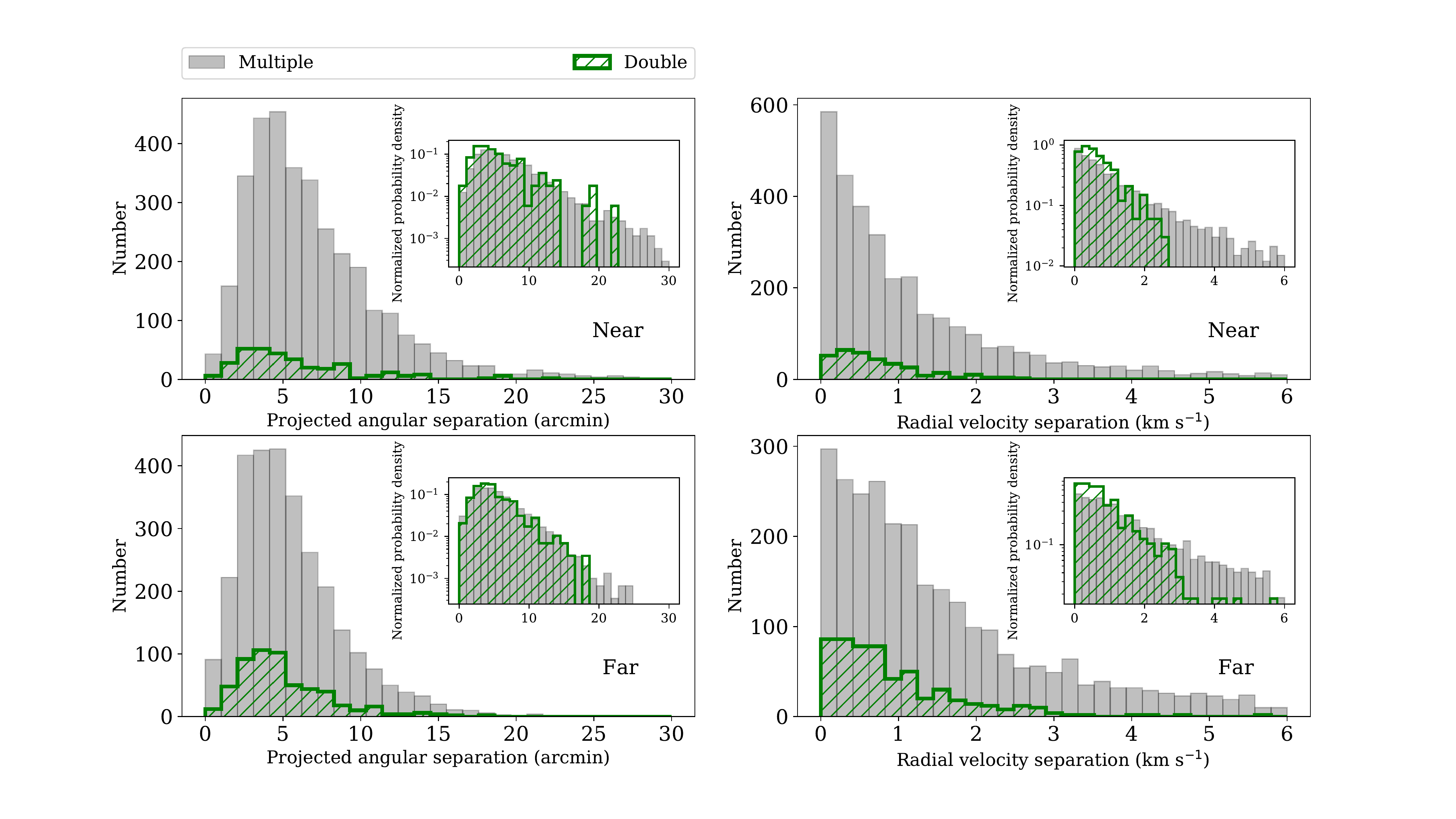} 
    \caption{The number distributions of projected angular separations and radial velocity separations 
    between the $^{13}$CO structures within $^{12}$CO clouds in the near (top) and far (bottom) groups, respectively. 
    The green and gray histograms represent the values from $^{13}$CO structures within 
    $^{12}$CO clouds having double and multiple $^{13}$CO structures, respectively. 
    In the up-right corner of each panel, the histograms present the corresponding 
    distribution of their normalized probability densities. The normalized probability 
    densities are presented as log scales, the separation values are binned as linear scales. \label{fig:fsepa}}
\end{figure*}

\begin{figure*}[ht]
    \plotone{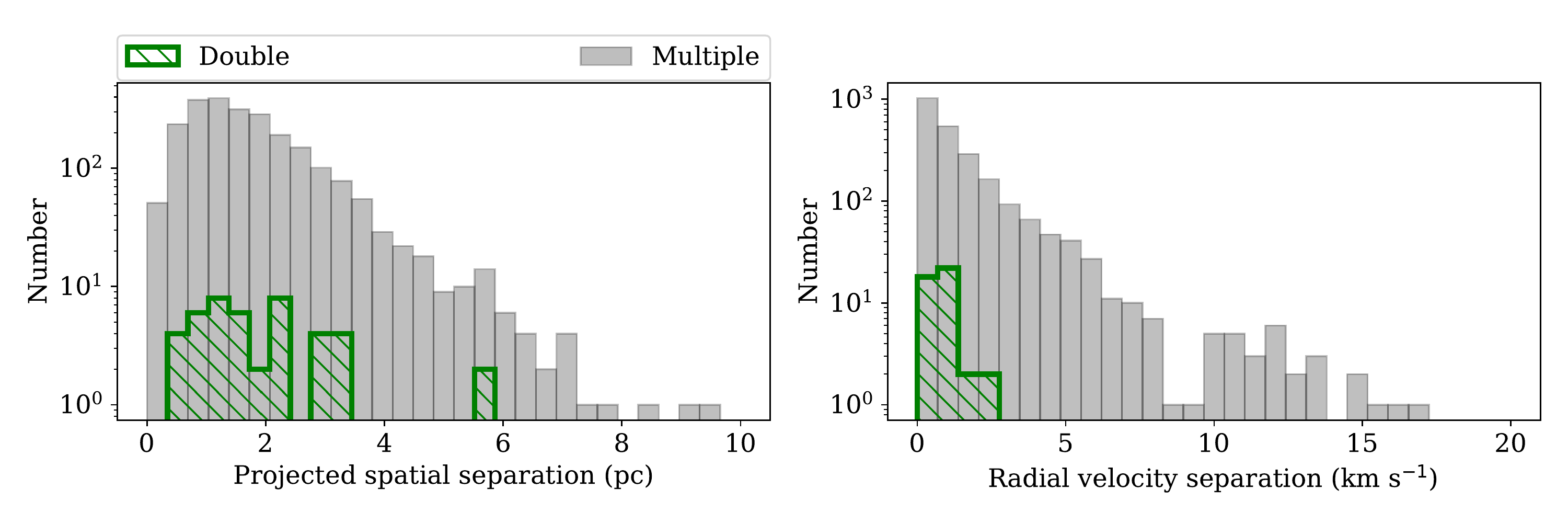} 
    \caption{The distributions of the linear separations and radial velocity separations 
    between the $^{13}$CO structures within 154 $^{12}$CO clouds. These $^{12}$CO clouds 
    have available distance information, which are measured with the background-eliminated 
    extinction-parallax method using extinctions and Gaia DR2 parallaxes in \cite{Yan2021b}. 
    The green and gray histograms represent the values from $^{13}$CO structures within $^{12}$CO clouds 
    having double and multiple $^{13}$CO structures, respectively. The number of $^{13}$CO 
    structures are presented as log scales, the separation values are binned as linear scales. \label{fig:fsepa_dist}}
\end{figure*}

\begin{figure*}[ht]
    \plotone{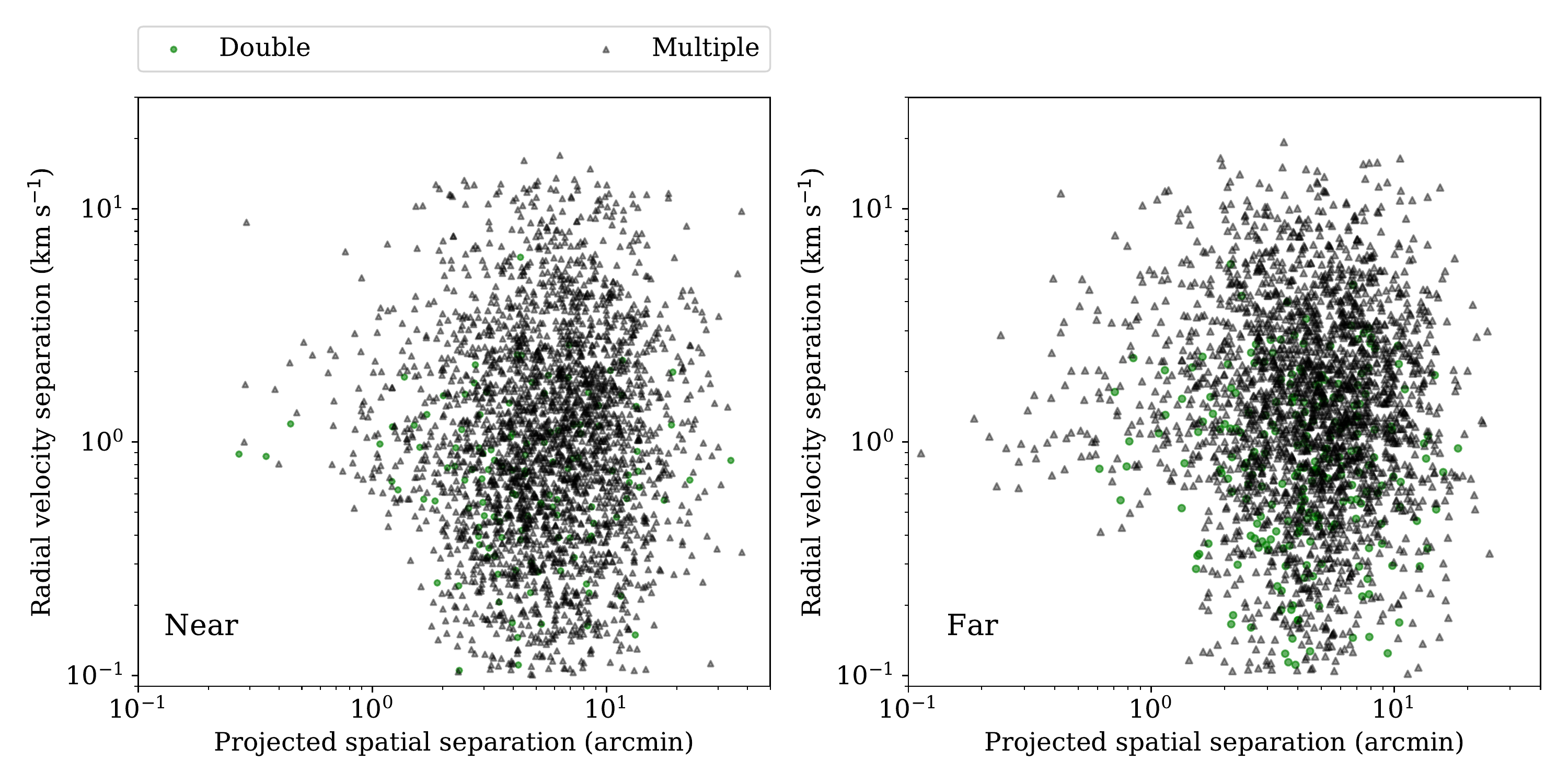} 
    \caption{The correlation between the projected angular separations and radial velocity separations 
    of the $^{13}$CO structures within $^{12}$CO clouds in the near and far groups, respectively. 
    The green and gray points represent the values from $^{13}$CO structures in the `double' 
    and `multiple' regime, respectively. \label{fig:fsepa_corr}}
\end{figure*}

\subsection{Projected angular separations and radial velocity separations}\label{sec:separation}

Since there can be more than one $^{13}$CO structure in a single $^{12}$CO cloud, how about the 
relative spatial distribution and kinematical motions between these $^{13}$CO structures? 
We compute the minimal spanning tree (MST) using the Kruskal algorithm in the PYTHON 
package\footnote[1]{https://docs.scipy.org/doc/scipy/reference/generated/scipy.sparse.csgraph.minimum\_spanning\_tree.html} 
to connect the whole $^{13}$CO structures in a single $^{12}$CO cloud together, 
while minimizing the total sum of the spacings between these $^{13}$CO structures. 
The observed angular separations between $^{13}$CO structures are the projection of 
their true 3-D spacings onto the plane of the sky,  
which are calculated as the euclidean distances between the central Galactic coordinates of $^{13}$CO structures. 
Their central Galactic coordinates are the averaged Galactic coordinate over all voxels within extracted 
$^{13}$CO structures, weighted by the corresponding $^{13}$CO(1-0) line intensities. 
The central velocity is the averaged radial velocity over all voxels within the extracted $^{13}$CO structure, 
weighted by the $^{13}$CO line intensities \citep{Rosolowsky2006}. 
The radial velocity separations are determined by the absolute differences 
between the central velocities of these connected $^{13}$CO structures.  
Figure \ref{fig:fMST} illustrates the computation of an MST over a $^{12}$CO cloud having four $^{13}$CO structures. 
Moreover, we calculate the projected angular separations and the radial velocity separations 
between $^{13}$CO structures along the MST in each $^{12}$CO cloud. 
For the $^{12}$CO clouds having double (443) or multiple (560) $^{13}$CO structures, 
the figures showing the distribution of their internal $^{13}$CO structures and the computed MSTs 
connecting the $^{13}$CO structures are available at DOI:\href{https://doi.org/10.57760/sciencedb.06653}{10.57760/sciencedb.06653}. 
In Figure \ref{fig:ftree} - \ref{fig:ftree1}, we present several samples. 
Moreover, the tables including the values of projected angular separations 
and radial velocity separations between the $^{13}$CO structures with each $^{12}$CO cloud are also 
published in the uniform DOI \citep{Yuan2022c}.

\begin{deluxetable*}{lcccccccc}
    \tablenum{1}
    \tablecaption{The projected spatial separations and radial velocity separations between $^{13}$CO structures within $^{12}$CO clouds \label{tab:t12co}}
    \tablewidth{0pt}
    \tablehead{\colhead{Types} & \colhead{Groups} &\colhead{Environment}  & \colhead{0.05} & \colhead{0.25} & \colhead{0.5} & \colhead{0.75} & \colhead{0.95} & \colhead{Mean} \\ 
    }
    \startdata
    \multirow{6}{*}{Spatial separation} & \multirow{2}{*}{Near (arcmin)} 
    & Double & 1.5 & 3.0 & 4.6 & 7.5 & 13.6 & 6.0 \\
     & & Multiple & 2.0 & 3.8 & 5.8 & 8.9 & 15.3 & 6.9 \\\cline{2-9}
    &\multirow{2}{*}{Far (arcmin)} & Double & 1.5 & 2.9 & 4.3 & 6.6 & 10.6 & 5.0 \\
     & & Multiple & 1.4 & 3.1 & 4.9 & 7.1 & 12.0 & 5.5 \\\cline{2-9}
    & \multirow{2}{*}{Subsample (pc)}
     & Double & 0.7 & 1.1 & 1.6 & 2.3 & 3.4 & 1.9 \\
     & & Multiple & 0.5 & 1.0 & 1.5 & 2.3 & 3.9 & 1.8 \\\hline
    \multirow{6}{*}{Velocity separation} & \multirow{2}{*}{Near (km s$^{-1}$)} 
    & Double & 0.05 & 0.28 & 0.57 & 0.95 & 1.93 & 0.75 \\
    & & Multiple & 0.06 & 0.32 & 0.8 & 1.8 & 5.62 & 1.52 \\\cline{2-9}
    &\multirow{2}{*}{Far (km s$^{-1}$)} & Double & 0.08 & 0.35 & 0.71 & 1.24 & 2.64 & 0.94 \\
    & & Multiple & 0.1 & 0.55 & 1.2 & 2.49 & 6.57 & 1.98 \\\cline{2-9}
    & \multirow{2}{*}{Subsample (km s$^{-1}$)} & Double & 0.11 & 0.28 & 0.76 & 0.95 & 1.9 & 0.75 \\
    & & Multiple & 0.06 & 0.33 & 0.83 & 1.77 & 4.96 & 1.44 \\\hline
    \enddata
    \tablecomments{The quantiles at 0.05, 0.25, 0.5, 0.75 and 0.95 for each parameter in its sequential data and its mean value. 
    The `Near’ and `Far’ represent for the MCs in the near and far groups, respectively. 
    The `Subsample' are the 154 MCs with available distance informations. 
    The `Double' and `Multiple' are corresponding to the MCs having double and multiple (more than two) $^{13}$CO structures, respectively. \label{tab:tsepa}}
\end{deluxetable*}

Considering the linear separations of $^{13}$CO structures are determined by their angular separations and distances, 
the resultant separations between $^{13}$CO structures are presented as near and far groups. 
Figure \ref{fig:fsepa} shows the distributions of the projected angular separations and radial velocity separations 
between the $^{13}$CO structures within each $^{12}$CO cloud in the near and far groups, respectively. 
To compare the distribution of separations in the multiple and double regimes, 
we also plot their normalized probability densities as insets in Figure \ref{fig:fsepa}. 
The normalized probability densities with a large dynamic range of several orders of magnitude 
are presented as log scales, the corresponding separation values having a narrow range are binned as linear scales.
Moreover, the quantiles at 0.05, 0.25, 0.5, 0.75, and 0.95 and the mean values of the projected angular separations 
and radial velocity separations are also listed in Table \ref{tab:tsepa}.  
For the angular separations, $\sim$ 50$\%$ of the values are in a narrow range of 3 -- 7 arcmin. 
We also find that the $^{13}$CO structures in the `multiple' regime have large separations ($>$ 20 arcmin), 
which are not observed in the `double' regime. 
Although the angular separations in the far group are slightly lower than those in the near group, 
the linear separations of $^{13}$CO structures in the far group are systematically higher than those in the near group 
by a factor of $\sim$ 3, considering the typically kinematical distances of $^{12}$CO clouds 
($\sim$ 0.5 kpc for the near group and $\sim$ 2.0 kpc for the far group). 
Under the uniform sensitivity and spatial angular resolution, due to the beam dilution effect, 
the isolated structures with smaller and lower line intensities may be diluted and further undetected, 
or several small structures are blended together as one structure \citep{Louvet2021}.
Hence the observed $^{12}$CO clouds and their internal $^{13}$CO structures 
in the far group tend to have relatively large spatial scales and high line intensities. 
Thus the spatial separations for $^{13}$CO structures in the near and far groups may reflect the spatial distribution of 
$^{13}$CO structures with different spatial scales.  

For a subsample with 154 $^{12}$CO clouds, which have the available distance information from \cite{Yan2021b}, 
the linear separations between their $^{13}$CO structures concentrate on a range of 1.0 -- 2.3 pc, their median and mean 
values are about 1.5 pc and 1.8 pc, respectively, as presented in Figure \ref{fig:fsepa_dist}. 
We should note that the MCs in this subsample tend to have large angular sizes, but small distances. 
$\sim$ 75$\%$ of their distances are less than or close to $\sim$ 1 kpc, 
mainly in the near group, as presented in Figure \ref{fig:fdistance}. 
Their $^{13}$CO structures in the `double' regime are also incomplete. 
The linear separations from this subsample may not completely represent the distribution of linear separations for the whole $^{13}$CO 
structures, but they still provide a reference for the preferred linear separation in the near group.

For the radial velocity separations between $^{13}$CO structures in each $^{12}$CO cloud, 
$\sim$ 50$\%$ of them are in a range of 0.3 -- 2.5 km s$^{-1}$, as listed in Table \ref{tab:tsepa}. 
The `multiple' regime also has large velocity separations ($\sim$ 3 -- 6 km s$^{-1}$), 
which are not nearly observed in the `double' regime. 
In addition, according to their quantiles listed in Table \ref{tab:tsepa}, 
we find the velocity separations in the far group are higher than those in the near group 
by a factor of $\sim$ 1.3.  

\begin{figure*}[ht]
    \plotone{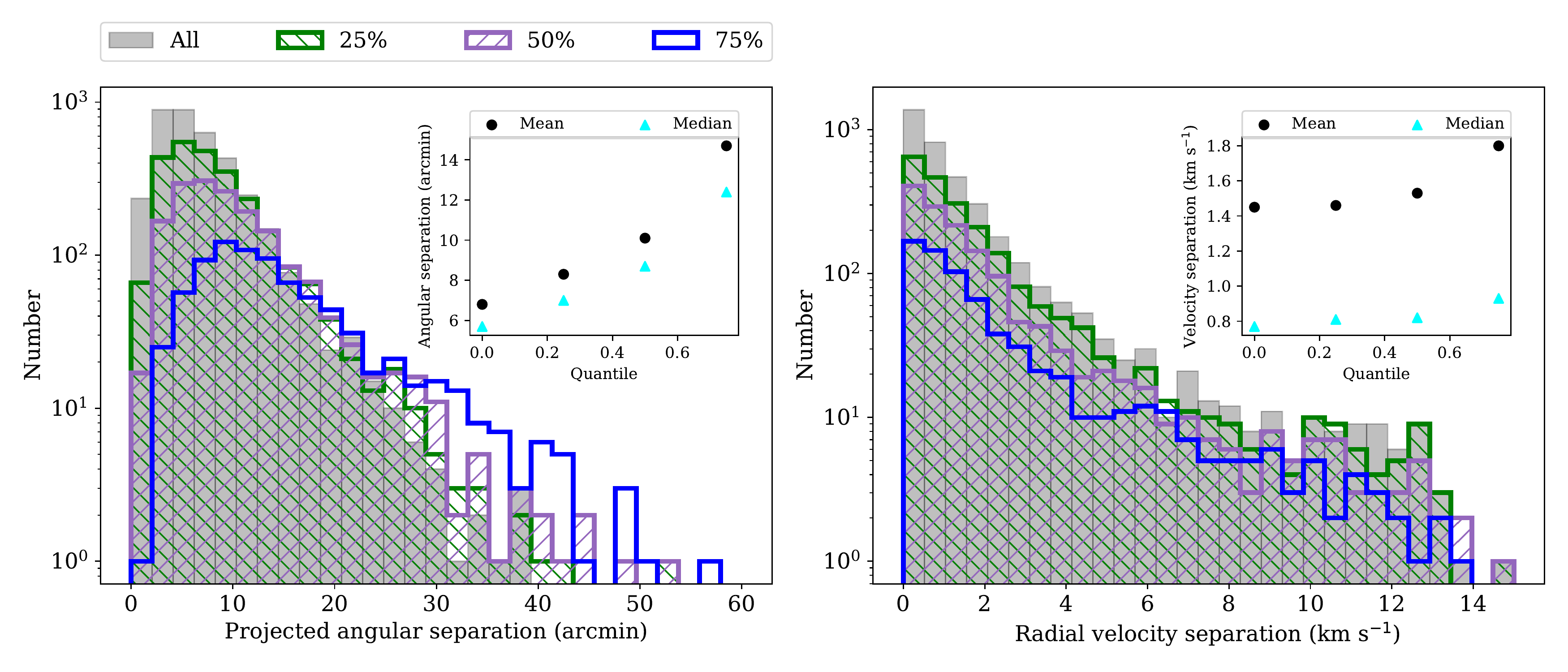} 
    \caption{The distribution of angular (left) and velocity (right) separations between the $^{13}$CO structures 
    with the different integrated fluxes. The gray histograms represent the whole $^{13}$CO structures in the near group. 
    The green, purple, and blue ones represent the near $^{13}$CO structures with integrated fluxes 
    larger than their quantile values at 0.25 (0.9 K km s$^{-1}$ arcmin$^{2}$), 
    0.5 (1.7 K km s$^{-1}$ arcmin$^{2}$), and 0.75 (4.6 K km s$^{-1}$ arcmin$^{2}$), respectively. 
    In the up-right corner of each panel, the scatter plot show the mean (black) and median (cyan) 
    values of angular (left) and velocity (right) separations for each sample. \label{fig:fsepa_cuts}}
\end{figure*}
Moreover, in Figure \ref{fig:fsepa_corr}, we present the correlation between 
the projected angular separations and radial velocity separations of $^{13}$CO structures in the near and far groups, 
respectively, the corresponding spearman's rank correlation coefficients is 0.04 for the near group and 
- 0.03 for the far group. That suggests there is no obvious correlation between the projected 
angular separations and radial velocity separations. 
Both angular separations and velocity separations concentrate on a certain range.  

Overall, the projected angular separations between $^{13}$CO structures, either in the near or far group, center on a 
range of $\sim$ 3 -- 7 arcmin. Their radial velocity separations mainly distribute from 0.3 km s$^{-1}$ to 2.5 km s$^{-1}$. 
The $^{13}$CO structures in the `multiple' regime exist large separations, either 
angular or velocity separations, which are not observed in the `double' regime.

\section{Discussion}
\subsection{Observational effects on the spatial separations of $^{13}$CO structures}
As mentioned in section \ref{sec:separation}, the linear separations among $^{13}$CO structures 
in the far group are $\sim$ 3 times those of $^{13}$CO structures in the near group. 
Under the uniform sensitivity and spatial angular resolution, 
the far $^{13}$CO structures with relatively small areas and low line intensities may not be observed, 
or several smaller structures are blended togehter \citep{Louvet2021}, due to the beam dilution effect. 
The differences between the linear separations are mainly caused by the observational effects 
or intrinsic distributions of $^{13}$CO structures. This needs a further analysis.

We take the whole $^{13}$CO structures in the near group as samples. 
These near $^{13}$CO structures are further grouped into three subsamples, 
whose $^{13}$CO integrated fluxes are larger than the quantile values 
at 0.25 (subsample-1), 0.5 (subsample-2), and 0.75 (subsample-3) of the $^{13}$CO integrated fluxes for the whole samples, respectively. 
The quantile values at 0.25, 0.5, and 0.75 are 0.9, 1.7, and 4.6 K km s$^{-1}$ arcmin$^{-2}$, respectively. 
Then we recompute the MST to determine the separations in each subsample.  
That is, the near $^{13}$CO structures with $^{13}$CO integrated fluxes less than the 0.9 K km s$^{-1}$ arcmin$^{-2}$ are removed, 
the rest are taken as the subsample-1 and further used to compute the MST for their parental $^{12}$CO clouds. 
The same process is taken for $^{13}$CO structures in subsample-2 and subsample-3. 

In figure \ref{fig:fsepa_cuts}, we present the distributions of the projected angular separations and 
radial velocity separations for the $^{13}$CO structures in the near samples, subsample-1, subsample-2, and subsample-3, 
respectively. We find that the mean and median angular separations gradually increase from the near sample to the subsamples. 
While their mean and median values of velocity separations do not vary significantly.  
The mean and median angular separations for the subsample-3 are about 2 times those in the whole near samples, 
while the factor is $\sim$ 1.2 for the radial velocity separations.

According to the fitted power-law distribution of $N \propto F^{-0.73}$ for the integrated fluxes in Figure \ref{fig:fnf}, 
we estimate $l_{A} \sim 1/\sqrt{N} \sim F^{0.4}$, where $l_{A}$ is the angular separation. 
This relation corresponds to the relation between the linear separations and the luminosities 
($l_{P} \sim \mathcal{L}^{0.4}$, where $l_{P}$ is the linear separation, $\mathcal{L}$ is the lumonosity), 
under the assumption of the close distances for $^{13}$CO structures in the near or far group. 
Thus the linear separations between $^{13}$CO structures depend on 
the number distribution of the observed $^{13}$CO structure luminosities. 
Under the noise rms of 0.25 K and the velocity resolution of 0.17 km s$^{-1}$ for $^{13}$CO line data, 
the integrated fluxes for the extracted $^{13}$CO structures by the DBSCAN algorithm are above $\sim$ 
0.5 K km s$^{-1}$ arcmin$^{2}$. 
For $^{13}$CO structures in the near and far groups, 
their kinematical distances concentrate on $\sim$ 0.5 kpc and $\sim$ 2 kpc, respectively. 
The limited luminosities for the near and far $^{13}$CO structures vary by a factor of $\sim$ 16. 
According to $l_{p} \sim \mathcal{L}^{0.4}$, 
the observed linear separations between $^{13}$CO structures in the near group are larger than 
those in the far group by a factor of 3 ($l_{\rm far}/l_{\rm near} \sim \mathcal{L}_{\rm far}^{0.4}/\mathcal{L}_{\rm near}^{0.4} \sim 3$), 
which is consistent with our observed results.
Thus we should note that the observed angular separations between $^{13}$CO structures depend on the number distribution of their integrated 
fluxes, which are limited by the spatial resolution and sensitivities of $^{13}$CO line data. 
Considering the observational effects, the whole $^{13}$CO structures are separated into near and far groups in the following analysis.

\begin{figure*}[h]
    \plotone{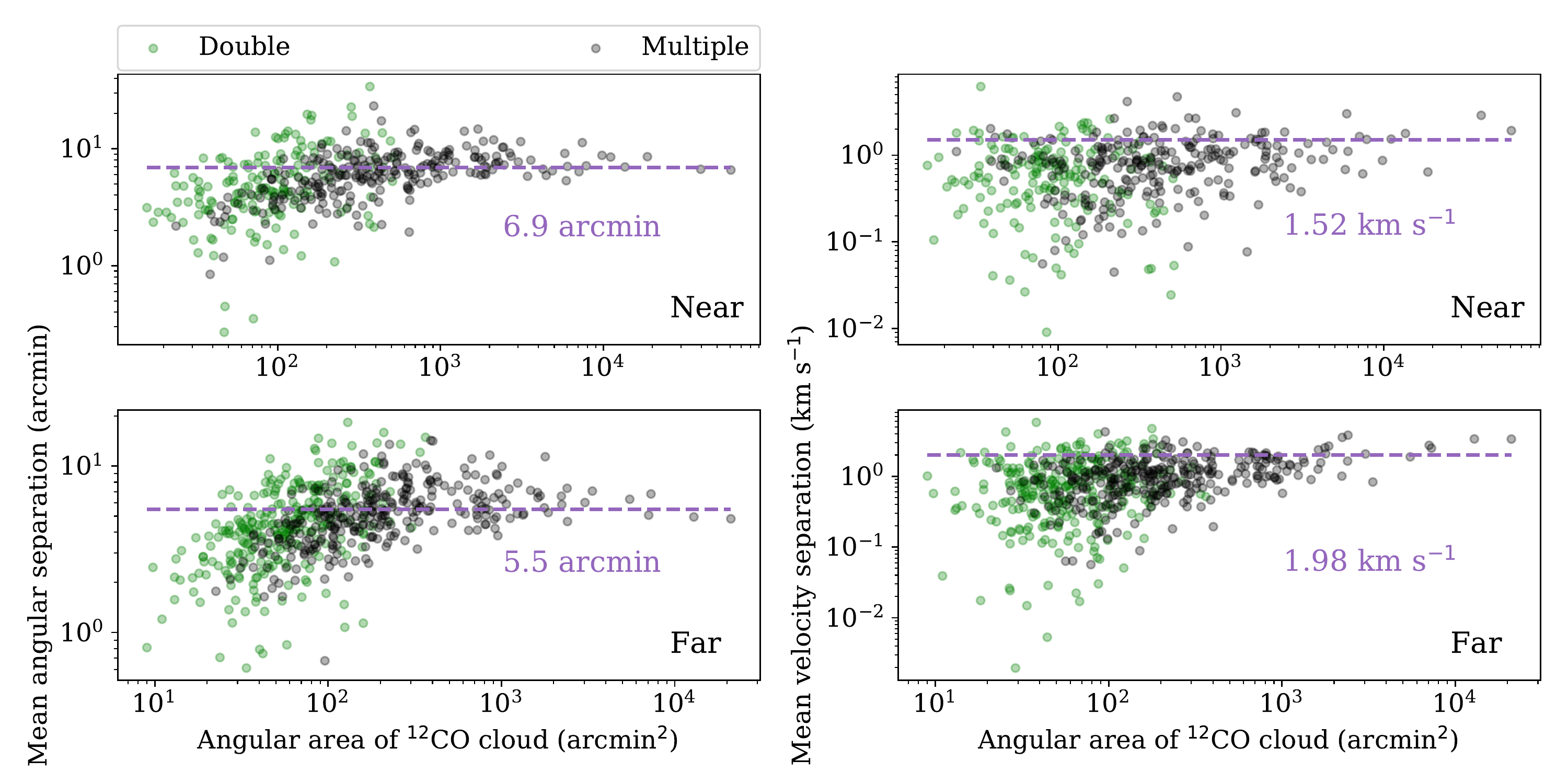} 
    \caption{Left panel: the correlations between the mean angular-separation of $^{13}$CO structures in each $^{12}$CO 
    cloud and the angular area of $^{12}$CO cloud in the near and far groups, respectively. 
    Each dot represents a single $^{12}$CO cloud, green and gray ones are for the $^{12}$CO clouds 
    having double and multiple $^{13}$CO structures, respectively.  
    The purple-dashed lines show the mean values for the whole angular separations in the 
    `multiple' regime, which are in the near (6.9 arcmin) and far (5.5 arcmin) groups, respectively. 
    Right panel: the correlations between the mean velocity-separations of $^{13}$CO structures in each $^{12}$CO cloud 
    and the angular area of $^{12}$CO cloud in the near and far groups, respectively. 
    The purple-dashed lines represent the mean values for the whole velocity separations 
    in the `multiple' regime, which are in the near (1.52 km s$^{-1}$) and far (1.98 km s$^{-1}$) groups, respectively. 
    \label{fig:fcorre_sepa}}
\end{figure*}

\begin{figure*}[h]
    \plotone{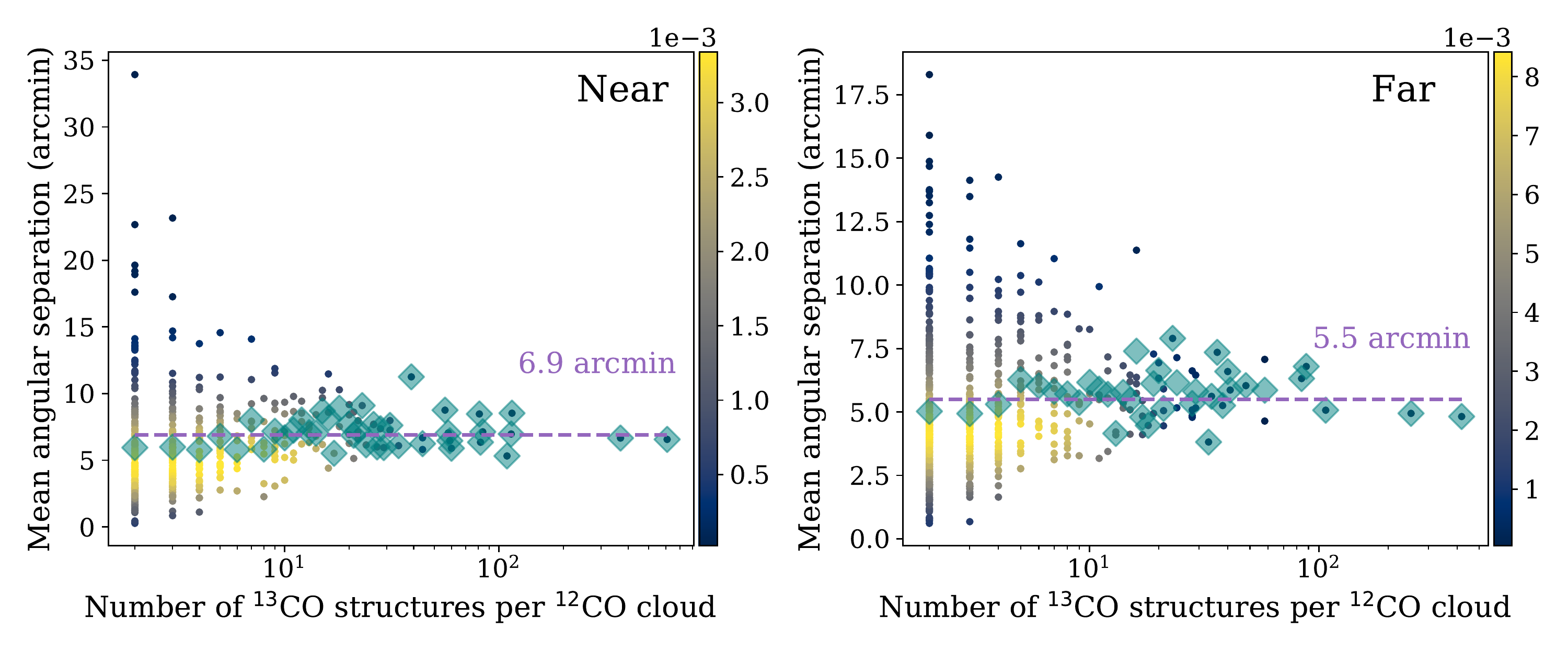} 
    \caption{The correlation between the mean angular-separation of $^{13}$CO structures within each $^{12}$CO cloud and 
    its harbored $^{13}$CO structure number in the near and far groups, respectively. 
    Each dot represents a single $^{12}$CO cloud.
    The colors on the dots represent the distribution of the probability density function of 
    the $^{12}$CO clouds (2D-PDF), which are calculated utilizing the kernel-density estimation through Gaussian kernels in the PYTHON package
    \href{https://docs.scipy.org/doc/scipy/reference/generated/scipy.stats.gaussian\_kde.html}{scipy.stats.gaussian\_kde}. 
    The teal diamonds represent the averaged values of the mean angular-separations in the $^{12}$CO clouds 
    having a uniform number of $^{13}$CO structures. The purple-dashed lines show the mean values for the whole angular separations in the 
    `multiple' regime, which are in the near (6.9 arcmin) and far (5.5 arcmin) groups, respectively. \label{fig:fcorre_mean}}
\end{figure*} 

\begin{figure*}[h]
    \plotone{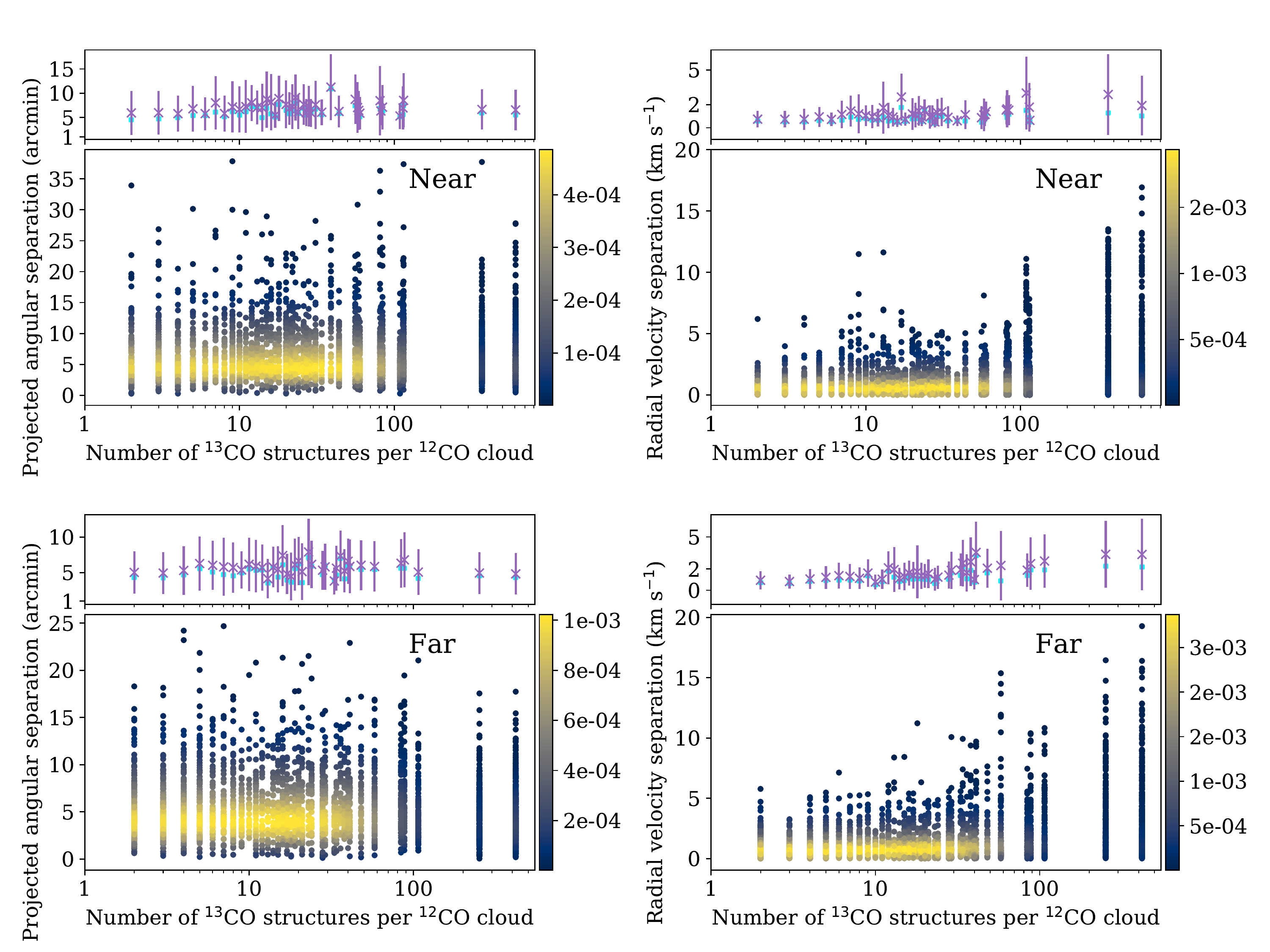} 
    \caption{The distribution of the whole projected angular separations and radial velocity separations 
    between $^{13}$CO structures within each $^{12}$CO cloud in the near and far groups, respectively. 
    These separations are grouped by the $^{13}$CO structure nummber in their natal $^{12}$CO clouds. 
    Each dot represents a value for the angular or velocity separation. 
    The colors on the dots represent the distribution of the probability density function of the separation values (2D-PDF). 
    In the top subpanel of each panel, the purple crosses and bars are the mean values and standard variances for 
    the whole separations in $^{12}$CO clouds having the same number of $^{13}$CO structures. 
    The cyan squares are the median values for those. \label{fig:fsepa_all}}
\end{figure*}

\subsection{Scaling relations between the $^{13}$CO structures and their natal $^{12}$CO clouds}

After presenting the distributions of the $^{13}$CO structure separations within each $^{12}$CO cloud, 
an interesting question concerns whether the relations between the $^{13}$CO structure separations
and their parental $^{12}$CO cloud scales exist.  

\subsubsection{A preferred angular separation between $^{13}$CO structures within $^{12}$CO clouds?}

Figure \ref{fig:fcorre_sepa} presents the correlation between the angular areas of the $^{12}$CO clouds 
and the mean angular and velocity separations of $^{13}$CO structures. 
We find that the mean angular separation of $^{13}$CO structures in each $^{12}$CO cloud 
is relatively dispersed in the small areas and gradually converge to a constant value 
($\sim$ 6.9 arcmin for the near group and $\sim$ 5.5 arcmin for the far group) as the increases of $^{12}$CO cloud areas. 
The constant values are the mean value of the whole angular separations in the `multiple' regime, 
shown as the purple-dashed lines. For the mean velocity separation in each $^{12}$CO cloud, 
there is a similar but slightly ascending trend along with the increase of $^{12}$CO cloud areas, 
their values concentrate in a narrow range of 0.3 - 2.5 km s$^{-1}$. 

Furthermore, we show the correlation between the mean angular separation and the $^{13}$CO 
structure number within each $^{12}$CO cloud in Figure \ref{fig:fcorre_mean}.  
There is a similar trend, i.e. the mean angular separation in each $^{12}$CO cloud gradually 
return to a nearly constant value with the increase of the internal $^{13}$CO structure number. 
We also average the mean angular separations in $^{12}$CO clouds harboring the same number of $^{13}$CO structures. 
These averaged values are also close to the constant value. 

It should be noted that the statistical errors for the mean separations in each $^{12}$CO cloud are different, 
due to that the number of $^{13}$CO structures within a single $^{12}$CO cloud ranges from 1 to $\sim$ 100. 
The mean separations in $^{12}$CO clouds with small scales ($\lesssim$ 10 $^{13}$CO structures) are scattered and 
those in the $^{12}$CO clouds with larger scales ($\gtrsim$ 10 $^{13}$CO structures) are more centered. 
Is this trend intrinsic or caused by the statistical errors? 
To investigate this question, we further present the distributions of the whole angular and velocity separations 
between $^{13}$CO structures in each $^{12}$CO cloud, 
grouped by the number of $^{13}$CO structure in their parental $^{12}$CO clouds. 
That is, the separations in the $^{12}$CO clouds harboring the same number of $^{13}$CO structures are 
put together, as shown in Figure \ref{fig:fsepa_all}. 
We find that these projected angular separations distribute in a concentrated range of $\sim$ 3 -- 7 arcmin, 
independent of what number of $^{13}$CO structures the $^{12}$CO clouds have. 
For each group of separations, which are in the $^{12}$CO clouds with the same number of $^{13}$CO structures, 
we also present their mean (purple crosses) and median (cyan squares) values, 
as well as their standard deviations (purple bars). 
These mean values are $\sim$ 7 arcmin for the near samples and $\sim$ 5.5 arcmin for the far samples, respectively. 
Their standard deviations are $\sim$ 4 arcmin for the near samples and $\sim$ 3 arcmin for the far samples.

The radial velocity separations are mainly in a narrow range of $\sim$ 0.3 -- 2.5 km s$^{-1}$. 
The mean values for velocity separations in each group are around $\sim$ 1.2 km s$^{-1}$ for the near samples and 
$\sim$ 1.7 km s$^{-1}$ for the far samples, 
also slightly increase to $\sim$ 2.5 km s$^{-1}$ in the $^{12}$CO clouds having large numbers of $^{13}$CO structures ($\gtrsim$ 80). 
In addition, their standard deviations also present a similar trend. 
One possibility for the increase of the radial velocity separation in the MCs with large scales 
is the star formation activities within dense structures. As presented in Figure \ref{fig:fstr_dist}, 
the $^{13}$CO structures with large angular areas, velocity spans, and integrated fluxes are 
observed in the `multiple' regime. 
Another possibility is due to the projection effect. The MST algorithm used to connect the 
$^{13}$CO structures minimizes the sum of their projected angular separations, 
not their radial velocity separations. 
The correspondence between the projected separations in angle and 
radial separations in velocity for these connected $^{13}$CO structures are not straightforward, 
especially for $^{12}$CO clouds having large numbers of $^{13}$CO structures.  

From above, we find the angular separations between $^{13}$CO structures in $^{12}$CO clouds, which are in either 
near or far groups, have nearly uniform mean values and standard deviations. 
That is, for the MCs at the same distance levels, which span a wide range of angular areas and the harbored $^{13}$CO structure counts, 
the spatial distributions of their internal $^{13}$CO structures are nearly consistent. 
The results also suggest there is a preferred angular separation between $^{13}$CO structures within these $^{12}$CO clouds.

\subsubsection{The development on the internal $^{13}$CO structure numbers of $^{12}$CO clouds}
\begin{figure*}[h]
    \plotone{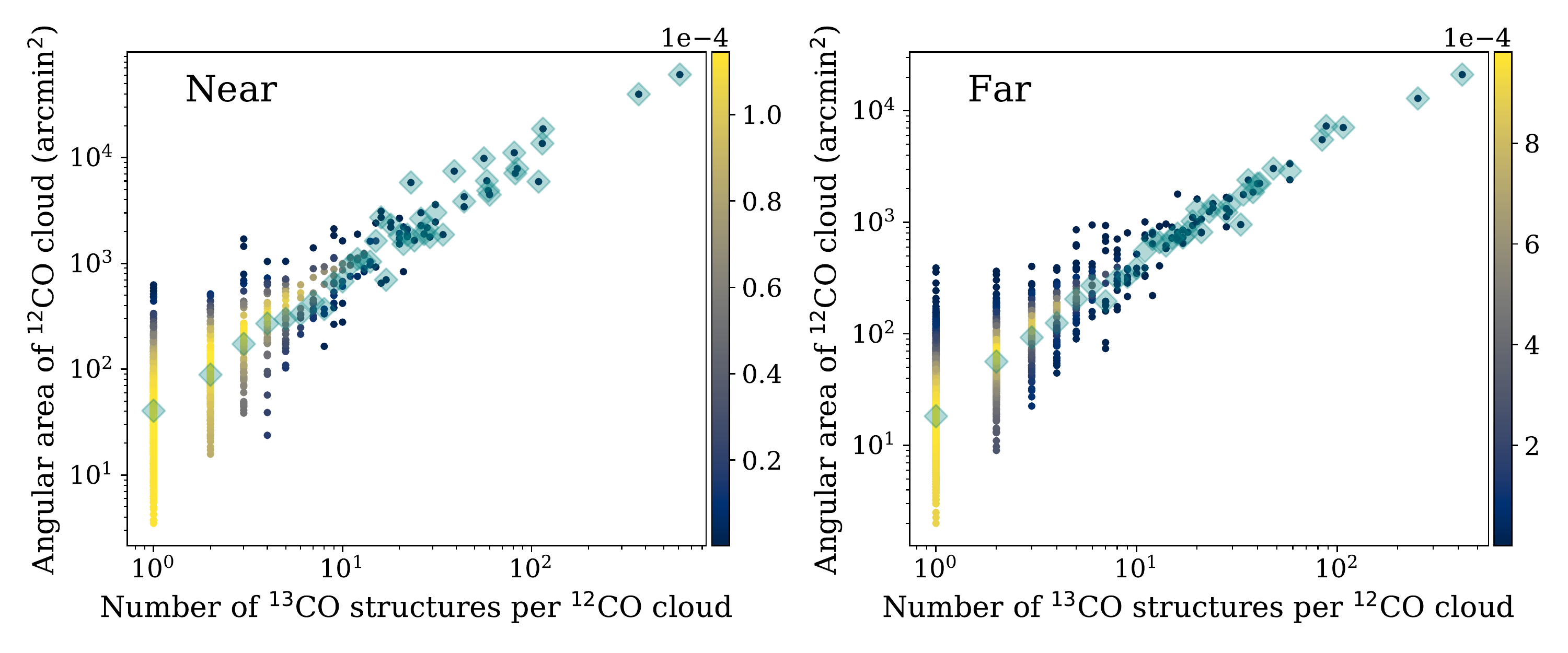} 
    \caption{The correlation between the $^{12}$CO cloud angular-areas and their harbored $^{13}$CO structure number 
    in the near and far groups, respectively.  
    Each dot represents a single $^{12}$CO cloud. 
    The colors on the dots represent the distribution of the probability density function of $^{12}$CO clouds (2D-PDF). 
    Teal diamonds represent the median values for the angular areas of $^{12}$CO clouds having the same number of $^{13}$CO structures. \label{fig:fcorre_counts}}
\end{figure*}

Since there is a preferred angular separation between $^{13}$CO structures within $^{12}$CO clouds,  
a relation between the spatial scale of $^{12}$CO cloud and its harbored $^{13}$CO structure number may exist. 
As presented in Figure \ref{fig:fcorre_counts}, we find a trend that 
the angular area of $^{12}$CO cloud increases with its harbored $^{13}$CO structures number, 
although it scatters in the regime for $^{12}$CO clouds with several $^{13}$CO structures. 
After taking the median angular areas for the $^{12}$CO clouds with the uniform number of $^{13}$CO structures, 
the trend traced by these median values is more obvious. 
It suggests that the more is their internal $^{13}$CO structures, 
the larger is the angular areas of $^{12}$CO clouds, 
under a preferred angular separation between $^{13}$CO structures.

\begin{figure*}[h]
    \plotone{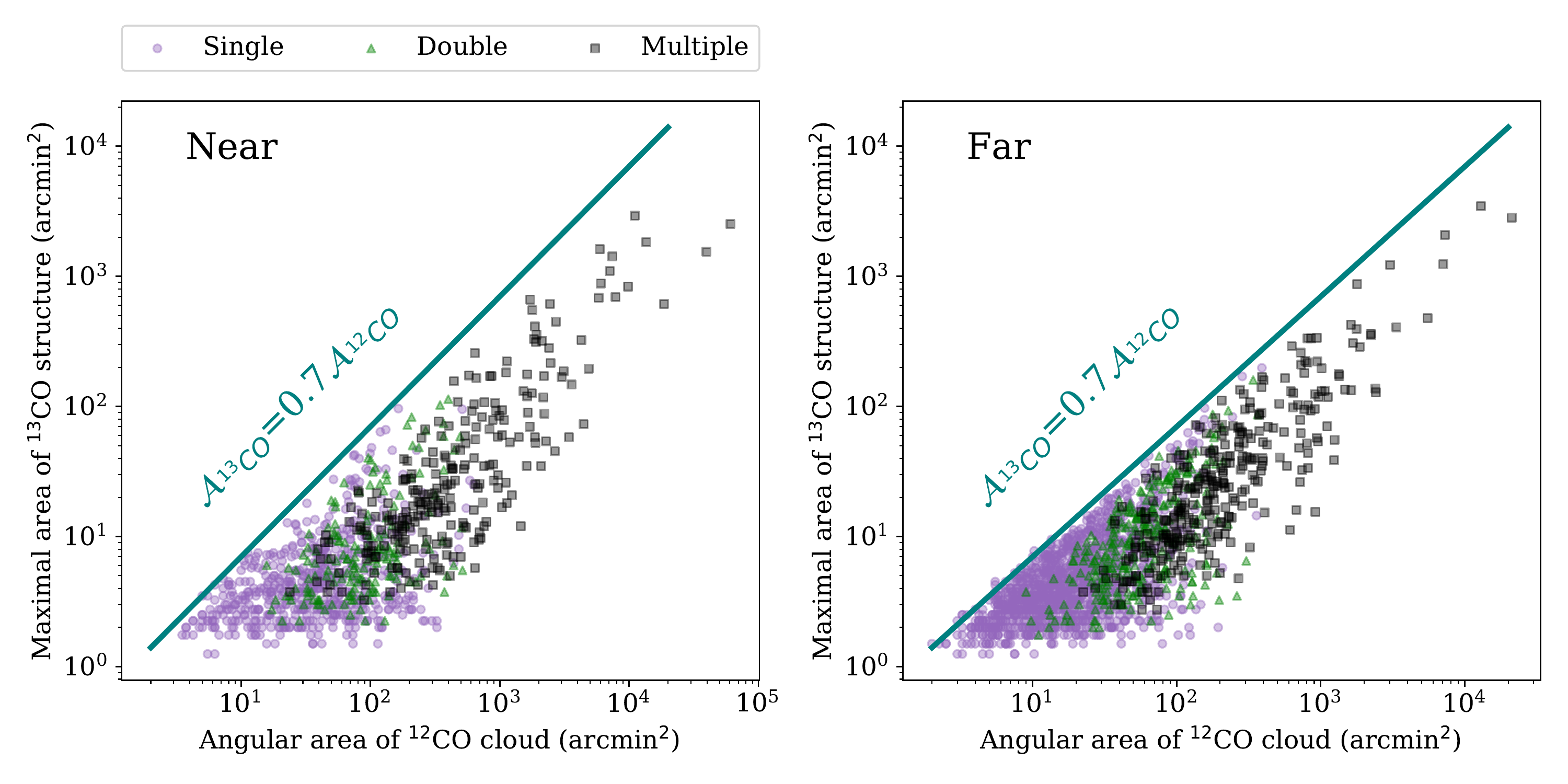} 
    \caption{The correlation between the maximum angular-area of of $^{13}$CO structures within a single $^{12}$CO cloud 
    and the angular area of its natal $^{12}$CO cloud. Each dot represents a single $^{12}$CO cloud, purple, green and gray ones are for the $^{12}$CO clouds 
    having single, double and multiple $^{13}$CO structures, respectively. 
    The teal lines show an upper limit that the whole $^{13}$CO emission area in a single MC generally do not exceed the 70$\%$ of 
    its $^{12}$CO emission area \citep{Yuan2022}. 
    \label{fig:fcorre_max}}
\end{figure*}

\subsubsection{The development on the maximum $^{13}$CO structure in each $^{12}$CO cloud \label{sec:max}} 
Since the internal $^{13}$CO structure count tend to increase with the $^{12}$CO cloud area, 
how about the area of a single $^{13}$CO structure? 
Do their values change with their parental $^{12}$CO cloud's areas? 
From the distribution of $^{13}$CO structure parameters in Figure \ref{fig:fstr_dist}, 
we find that the $^{13}$CO structures in the `multiple' regime have larger values in the angular 
areas, velocity spans, and integrated fluxes, which are not observed in the `single' and `double' regimes.
Here we further investigate the correlation between the maximal angular-area for $^{13}$CO structures in each 
$^{12}$CO cloud and the $^{12}$CO cloud angular-area. 
 
As presented in Figure \ref{fig:fcorre_max}, 
there is a roughly positive correlation between the maximal angular-area of $^{13}$CO structures in 
each $^{12}$CO cloud and the $^{12}$CO cloud angular-area, with the spearman's rank correlation coefficients of 0.7. 
The teal line shows an upper limit that the whole $^{13}$CO emission area in a single $^{12}$CO cloud 
generally do not exceed the 70$\%$ of its $^{12}$CO emission area \citep{Yuan2022}. 
The maximal $^{13}$CO structures in the `double' and `multiple' systems are lower than the upper limit, 
but still have a growing tendency among their areas and their parent $^{12}$CO cloud areas. 
Combining the distribution of $^{13}$CO structure parameters in the `multiple' regime in Figure \ref{fig:fstr_dist}, 
we find the $^{12}$CO clouds having larger scales tend to have $^{13}$CO structures distributed in the larger scales. 
This result suggests the development of large-scale $^{13}$CO structures is synchronized with the spatial scales 
of $^{12}$CO clouds.

\subsection{Implications of molecular clouds assembly and destruction}  

Since the spatial distributions of $^{13}$CO structures within $^{12}$CO molecular clouds 
spanning a wide range in sizes and environmental conditions are resembled, 
in addition, there is a preferred angular separation between $^{13}$CO structures.
These results imply the spatial distribution of internal $^{13}$CO structures in molecular clouds 
are affected by the universal physical mechanism.

The spatial distribution of substructures in molecular clouds, 
like the dense cores \citep{Andre2014, Tafalla2015, Kainulainen2017, Konyves2020}, 
young stars \citep{Gomez1993, Larson1995, Kraus2008}, and young stellar clusters \citep{Grasha2015} 
have been investigated. The observed core separations in filaments are interpreted by the two-mode filament 
fragmentation: for filaments in which the gravitational energy is dominated over turbulent, 
the ``cylindrical" fragmentations occur corresponding to clumps with a separation scale 
of $\sim$ 4 $\times$ filament width \citep{Inutsuka1992}, 
then the small-scale fragmentations at the effective Jeans length ($\lesssim$ 0.1 pc) follow \citep{Clarke2017, Konyves2020, Pineda2022}. 
For young stars, their spatial distributions exhibit self-similar or fractal clustering on the larger scales, 
and have a clear break from the self-similarity at a small scale, comparable to the Jeans length in typical molecular cloud cores 
\citep{Larson1995}. Thus, these studies suggest that gravitational instability is responsible for 
the spacings of small-scale and dense substructures in MCs, like clumps, cores, and young stars. 
For the MCs with distances information, the observed separations between $^{13}$CO structures in MCs are 
mainly in 1.0 pc -- 2.3 pc with a median of 1.5 pc, as listed in Table \ref{tab:tsepa}. 
Under consideration of the densities n $\sim$ 10$^{2}$ -- 10$^{3}$ cm$^{-3}$ and 
T $\sim$ 10 K for the MCs, the estimated Jeans length is in the range of $\sim$ 1.4 pc -- $\sim$ 0.4 pc. 
Our observed separations between $^{13}$CO structures are systematically larger than the Jeans length scales. 
This implies the regular separations of $^{13}$CO structures within $^{12}$CO clouds 
are hardly interpreted by the Jeans fragmentation.

In another picture, MCs appear to be transient objects that may 
never reach an equilibrium configuration \citep{Vazquez1995, Passot1995, Ballesteros-Paredes1999, Vazquez-Semadeni2003, Clark2012, Ballesteros2020}. 
They can form in the diffuse ISM through local compressions induced by the converging flows 
from the differential rotation and shear of the spiral arms \citep{Kwan1983, Tomisaka1984, Kwan1987}, 
the passage of HII regions or supernova shells \citep{Heiles1979, Heiles1984, Tenorio-Tagle1988, Hartmann2001}, 
shock-wave passage \citep{Bergin2004, Clark2012, Guo2021},
large-scale gravitational instability of rotating disk \citep{Lin1964, Roberts1969, Shetty2008, Tasker2009, Dobbs2011}, 
clouds agglomeration \citep{Oort1954, Field1965, Kwan1983, Tomisaka1984, Dobbs2014}, 
and cloud-cloud collisions \citep{Gilden1984, Fukui2014, Fukui2016, Dobbs2015, Gong2017}. 
In addition, the density fluctuations within MCs can be produced by a common hierarchy of interstellar converging flows, 
i.e. the supersonic dissipation in the ambient medium \citep{Larson1981, Myers1983, Falgarone1992, Cernicharo1985, Heyer2004, Falgarone2009}. 
An evidence generally used to support this picture is 
the results of the probability density functions of the molecular hydrogen column density (N-PDFs) of MCs 
\citep{Vazquez-Semadeni1994, Padoan1997, Scalo1998, Klessen2000}. 
\cite{Ma2021, Ma2022} has analyzed the N-PDFs for a sample of MCs using $^{13}$CO lines, 
and found that $\sim$ 72$\%$ of them present a log-normal distribution and 18$\%$ are fitted by a log-normal plus power-law function. 
However, none of the transient models provide the prediction for the regular separations 
between $^{13}$CO structures within $^{12}$CO clouds so far. 
Our observed results will provide a constraint for the transient picture of MCs. 
Most likely, the preferred separation will provide a fundamental unit for the assembly and destruction of MCs. 

From the scaling relations between the $^{12}$CO clouds and their internal $^{13}$CO structures, 
the spatial scales of $^{12}$CO clouds increase with their harbored $^{13}$CO structures number, 
under a preferred angular separation between $^{13}$CO structures. 
Among our 2851 MC samples, $\sim$ 60$\%$ harbor one single $^{13}$CO structure, $\sim$ 15$\%$ have 
double $^{13}$CO structures, the rest $\sim$ 20$\%$ have multiple (more than two) $^{13}$CO structures, 
as presented in Figure 11 in \cite{Yuan2022}. The mean and spread of the angular separations 
between $^{13}$CO structures within these $^{12}$CO clouds are resembled, independent of the $^{12}$CO 
cloud scales. According to these results, we propose an alternative picture for the assembly and destruction of molecular clouds: 
there is a fundamental separation between the internal structures of molecular clouds, 
hence the build-up or destruction of molecular clouds proceeds under this fundamental unit, 
although the origin of this fundamental separation is still needed to 
be identified. Moreover, the $^{12}$CO clouds having larger scales tend to have $^{13}$CO structures 
with larger scales, as mentioned in Section \ref{sec:max}.
That suggests the development of large-scale $^{13}$CO structures is synchronized with the $^{12}$CO cloud scales. 
Follow-up analysis on the velocity structures of MCs is essential for us to further verify this picture. 
The transient configuration and fundamental separations between high-density substructures 
can regulate the mass agglomeration in MCs, and further prevent higher star formation efficiency. 
Therefore the picture of fundamental units provides a natural explanation for the low star formation rate in MCs.

\section{Conclusions}
We use a sample of 2851 $^{12}$CO molecular clouds, inside which a total of 9566 $^{13}$CO gas structures are identified, 
to systematically investigate the spatial distribution and relative motion of $^{13}$CO structures within $^{12}$CO molecular clouds. 
The main conclusions are as follows: 

1. The projected angular separations center on a range of $\sim$ 3 -- 7 arcmin with a median of $\sim$ 5 arcmin, 
and the radial velocity separations mainly distribute from $\sim$ 0.3 km s$^{-1}$ to 2.5 km s$^{-1}$. 
The observed angular separations depend on the integrated fluxes spectra of the observed $^{13}$CO structures, 
which is limited by the spatial resolution and sensitivities of $^{13}$CO line data. 

2. The angular separations between $^{13}$CO structures within $^{12}$CO clouds have a preferred value, 
independent of the $^{12}$CO cloud angular areas and the numbers of their internal $^{13}$CO structure.  

3. A scaling relation showing the $^{12}$CO cloud area increases with its internal $^{13}$CO structure count is revealed. 
Moreover, it is found that the maximum angular area of $^{13}$CO structures in each $^{12}$CO cloud 
tends to increase with the angular area of $^{12}$CO cloud itself.     

\begin{acknowledgments}
    This research made use of the data from the Milky Way Imaging Scroll Painting (MWISP) project, 
    which is a multi-line survey in $^{12}$CO/$^{13}$CO/C$^{18}$O along the northern galactic plane with PMO-13.7m telescope. 
    We are grateful to all of the members of the MWISP working group, particulaly to the staff members at the PMO-13.7m telescope, 
    for their long-term support. This work was supported by the National Natural Science Foundation of China through grant 12041305. 
    F. Du is also supported by NNSFC through grant 11873094. 
    MWISP was sponsored by the National Key R\&D Program of China with grant 2017YFA0402701 
    and the CAS Key Research Program of Frontier Sciences with grant QYZDJ-SSW-SLH047. 
\end{acknowledgments}

\software{Astropy \citep{astropy2013, astropy2018}, Scipy \citep{scipy2020}, Matplotlib \citep{Hunter2007}}
    
\clearpage
\appendix
\renewcommand{\thefigure}{\Alph{section}\arabic{figure}}
\renewcommand{\theHfigure}{\Alph{section}\arabic{figure}}
\setcounter{figure}{0}

\section{Examples on the minimal spanning tree over $^{12}$CO clouds}
Figure \ref{fig:ftree} - \ref{fig:ftree1} present the $^{13}$CO structures within several $^{12}$CO molecular clouds and 
the calculated MSTs connecting the $^{13}$CO structures.
\begin{figure*}
    \plotone{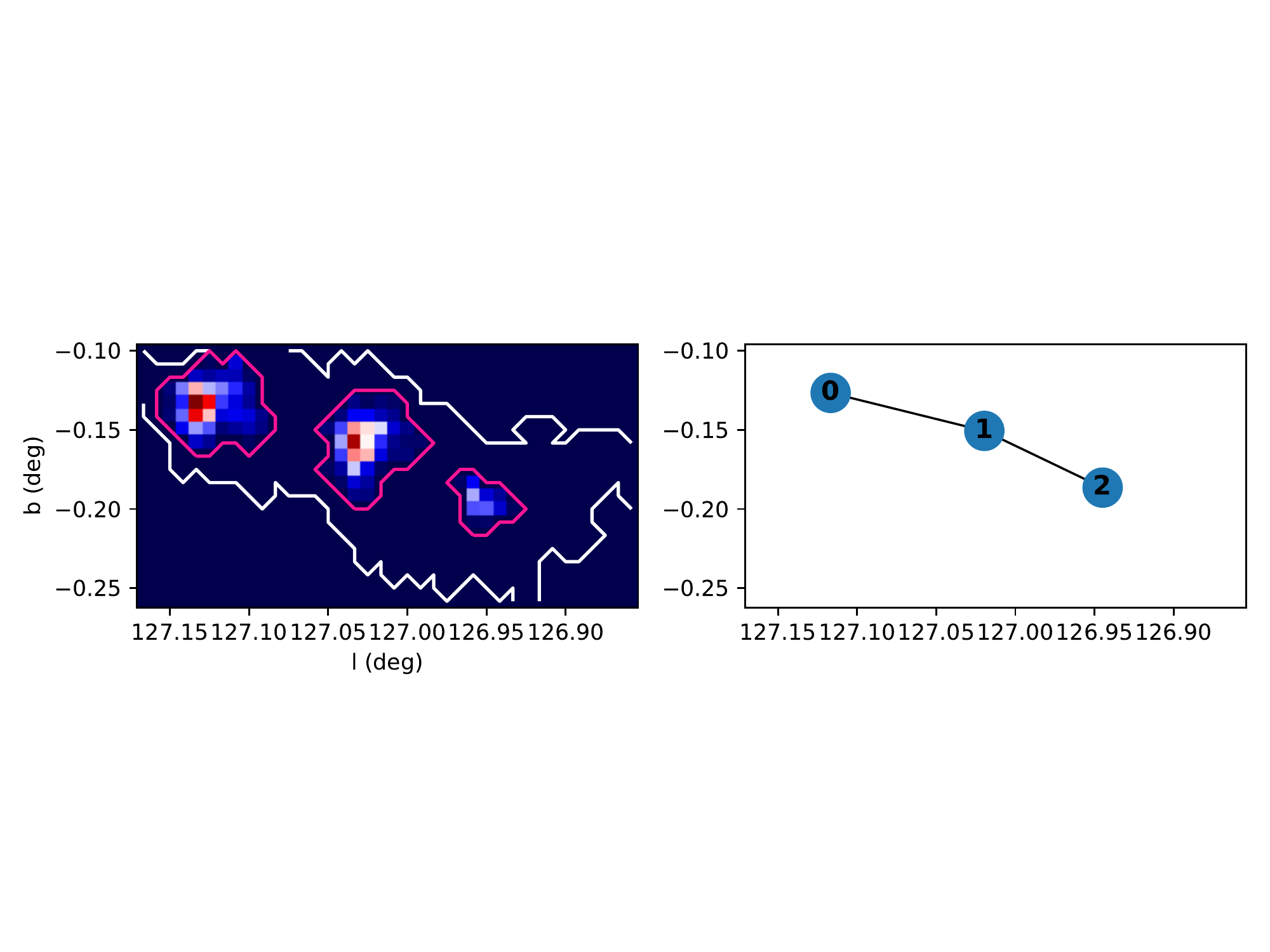}
    \caption{Left panel: the colormap represents the distribution of the velocity-integrated intensities of 
    $^{13}$CO line emission within a $^{12}$CO cloud. The white and magenta contours indicate the boundaries of the $^{12}$CO cloud and $^{13}$CO 
    structures, which are identified by the DBSCAN algorithm. 
    Right panel: the minimal spanning tree connects the internal $^{13}$CO structures over 
    a $^{12}$CO cloud. Each node sites the central Galactic coordinates for a $^{13}$CO structure. \label{fig:ftree}}
\end{figure*}

\begin{figure*}
    \plotone{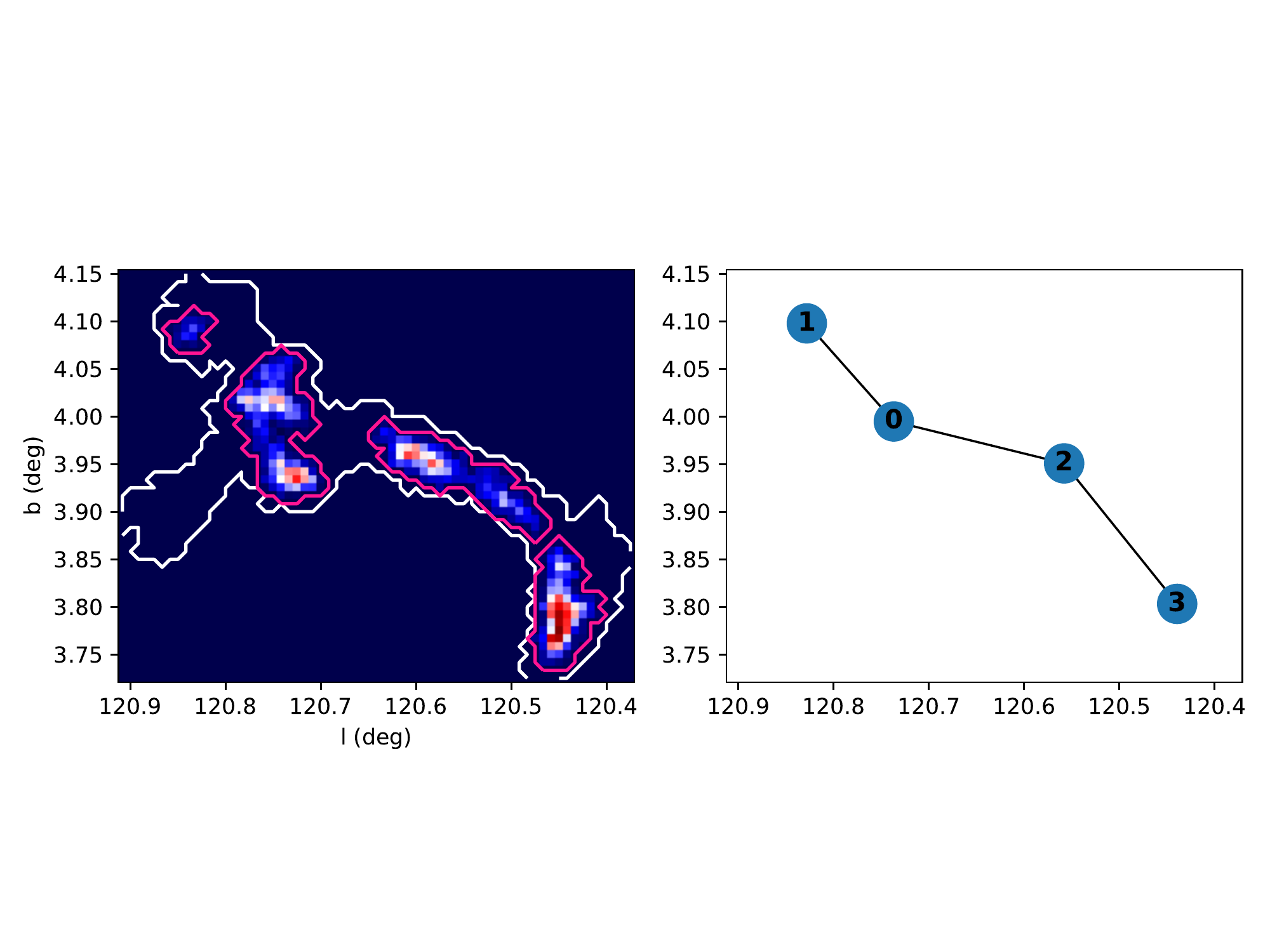}
    \caption{Continued from Fig. \ref{fig:ftree}.}
\end{figure*}

\begin{figure*}
    \plotone{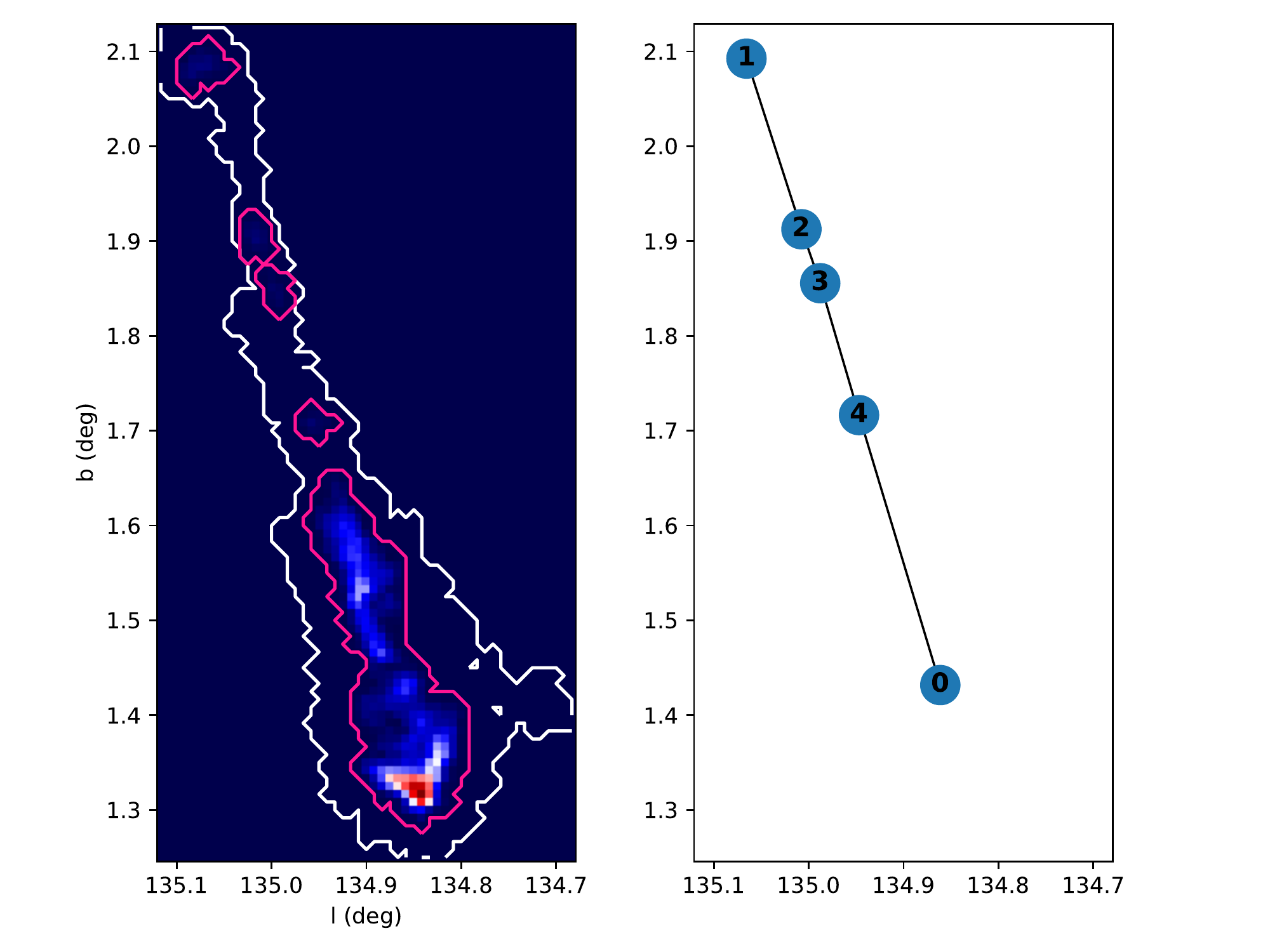}
    \caption{Continued from Fig. \ref{fig:ftree}.}
\end{figure*}

\begin{figure}
    \plotone{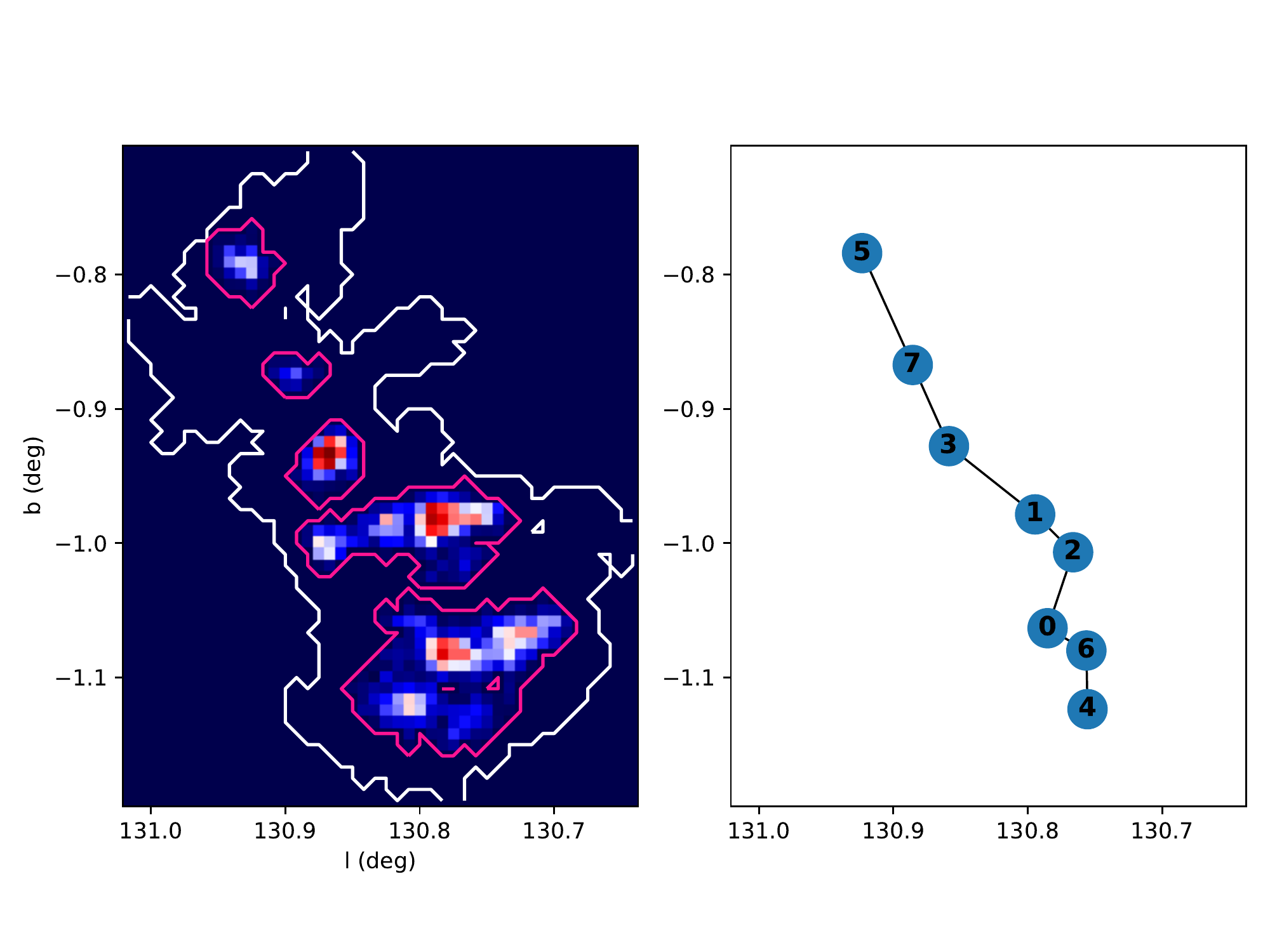}
    \caption{Continued from Fig. \ref{fig:ftree}.}
\end{figure}

\begin{figure}
    \plotone{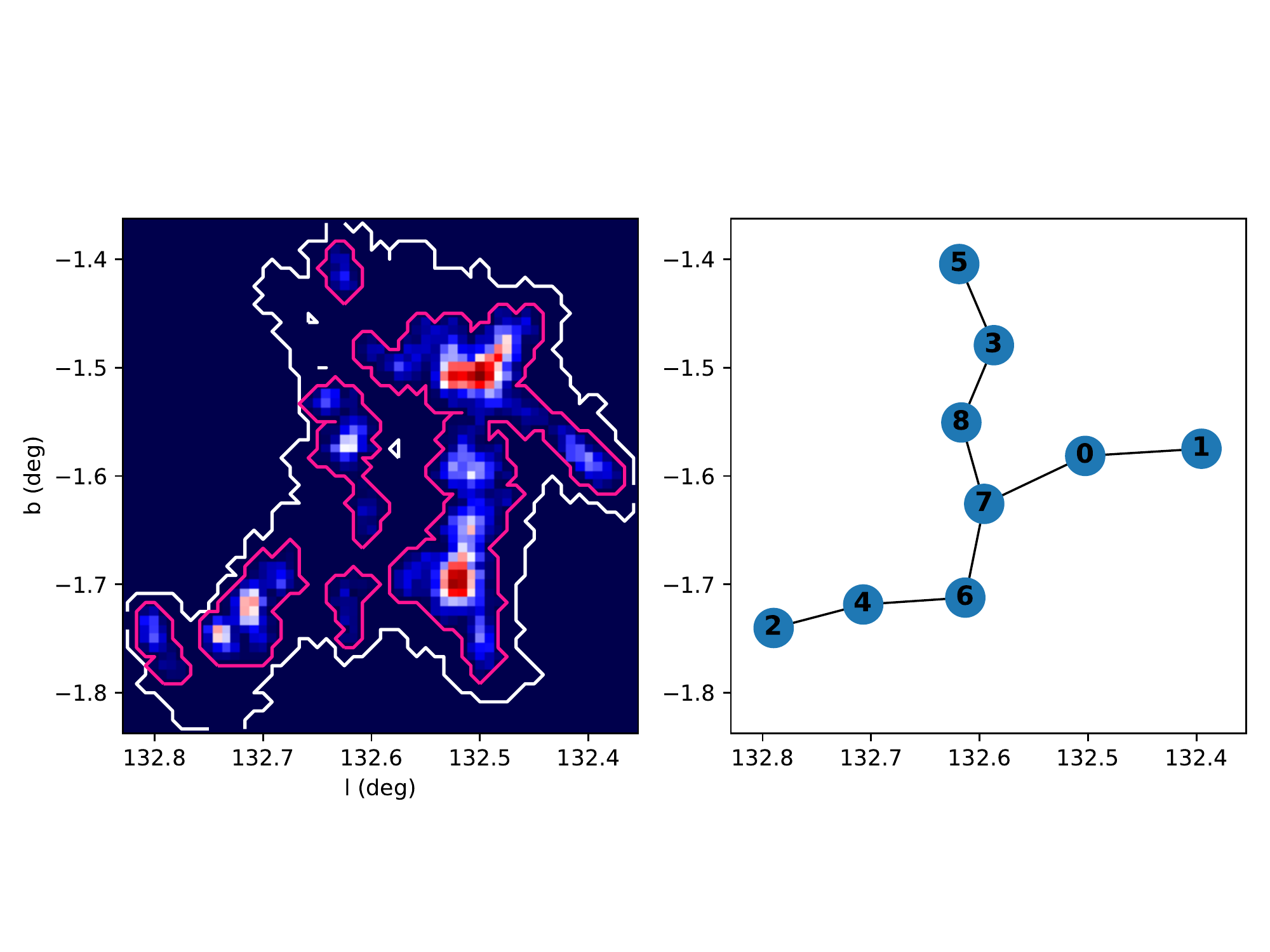}
    \caption{Continued from Fig. \ref{fig:ftree}.}
\end{figure} 

\begin{figure}
    \plotone{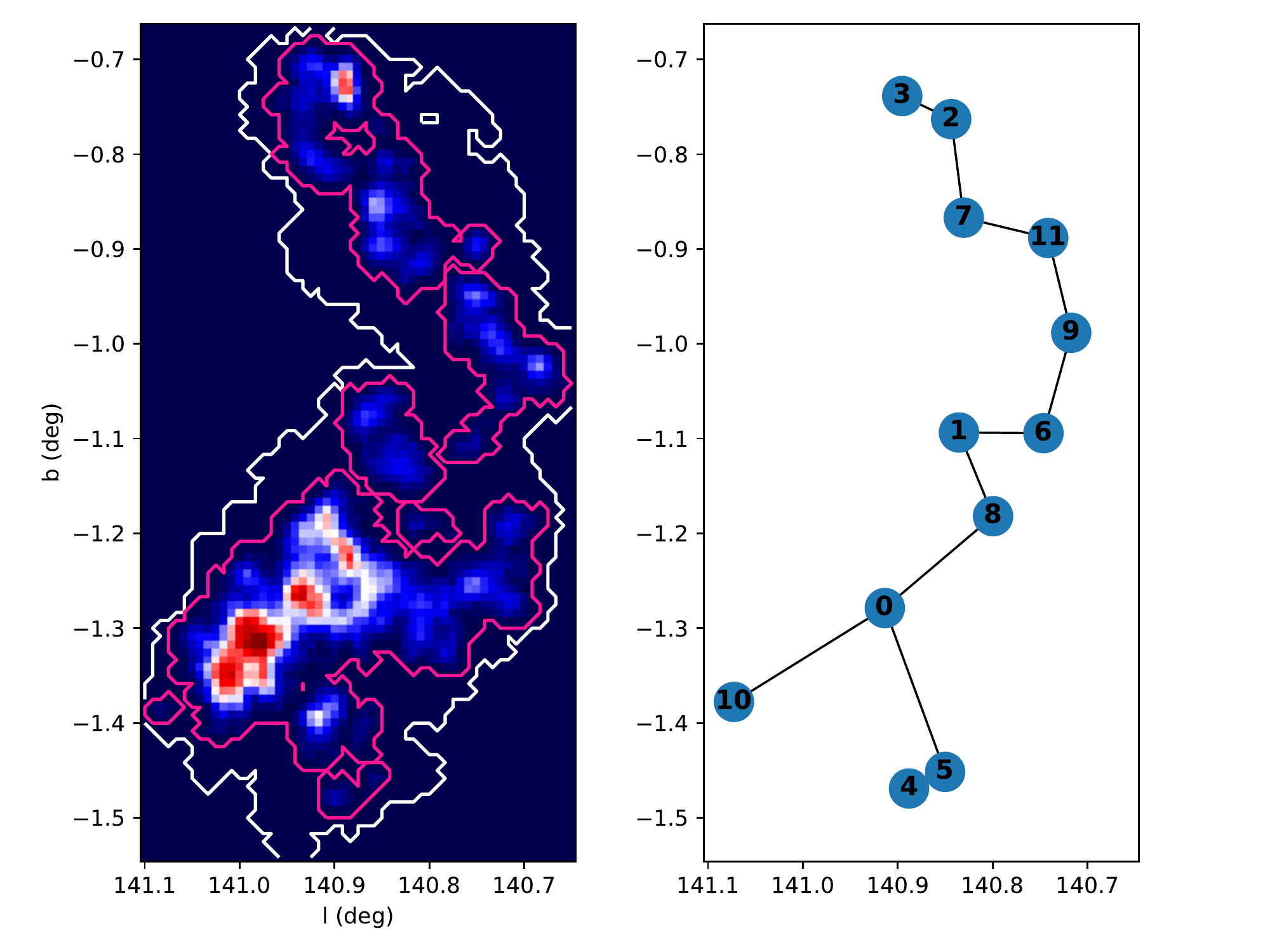}
    \caption{Continued from Fig. \ref{fig:ftree}.}
\end{figure}

\begin{figure}
    \plotone{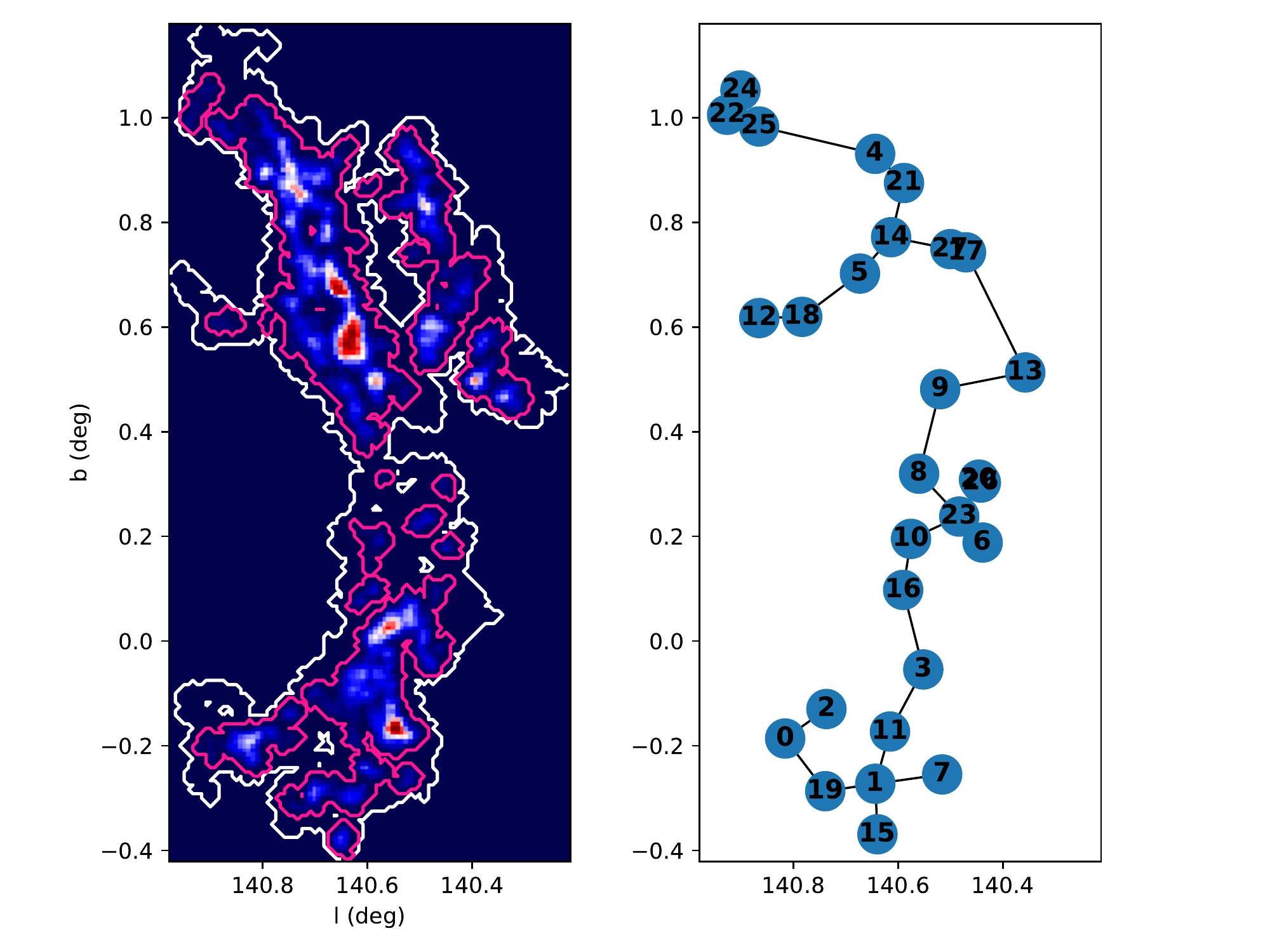}
    \caption{Continued from Fig. \ref{fig:ftree}.}
\end{figure}
\begin{figure}
    \plotone{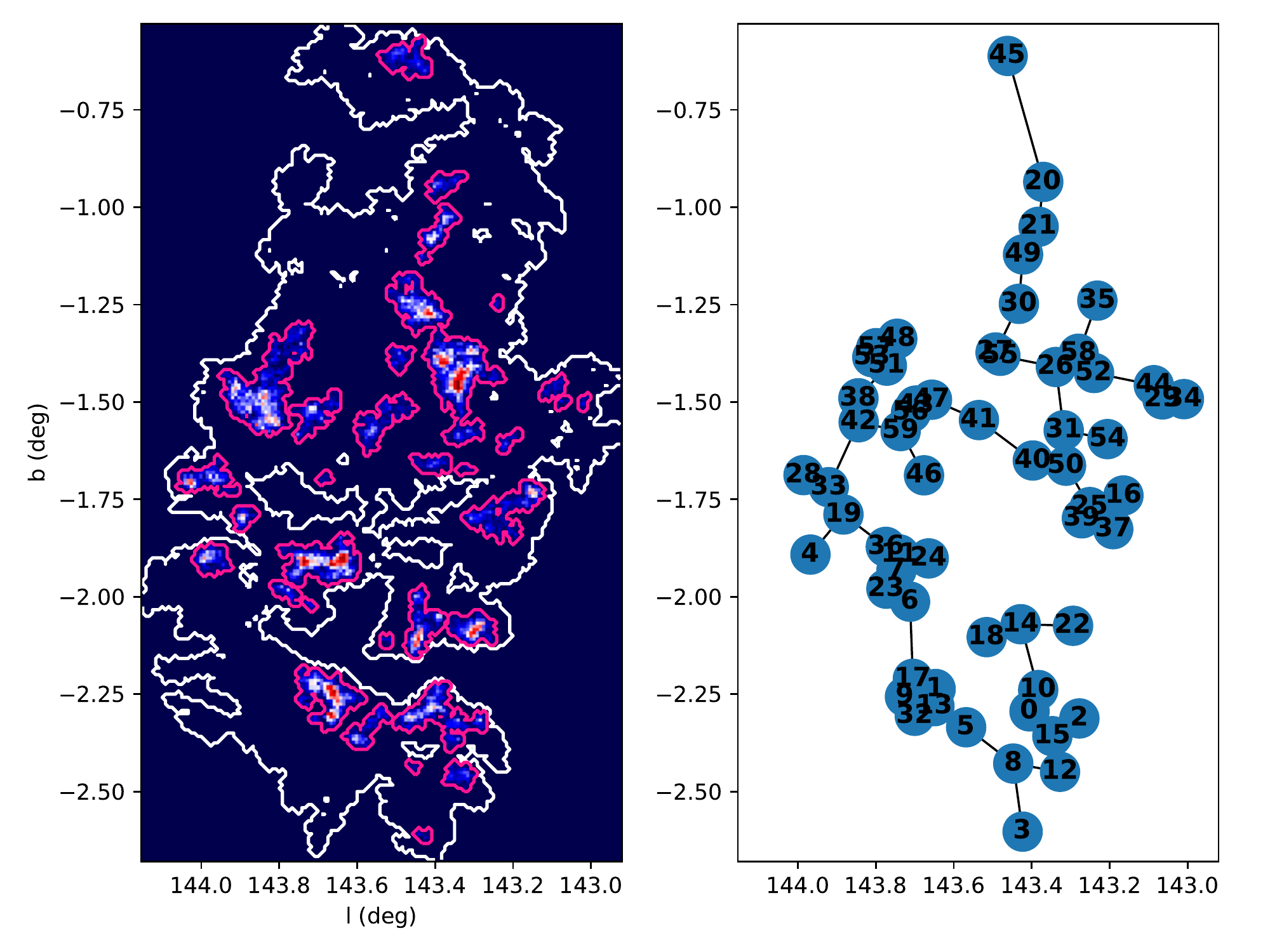}
    \caption{Continued from Fig. \ref{fig:ftree}. \label{fig:ftree1}}
\end{figure}

\section{The distances for 154 molecular clouds} 
\begin{figure}
    \plotone{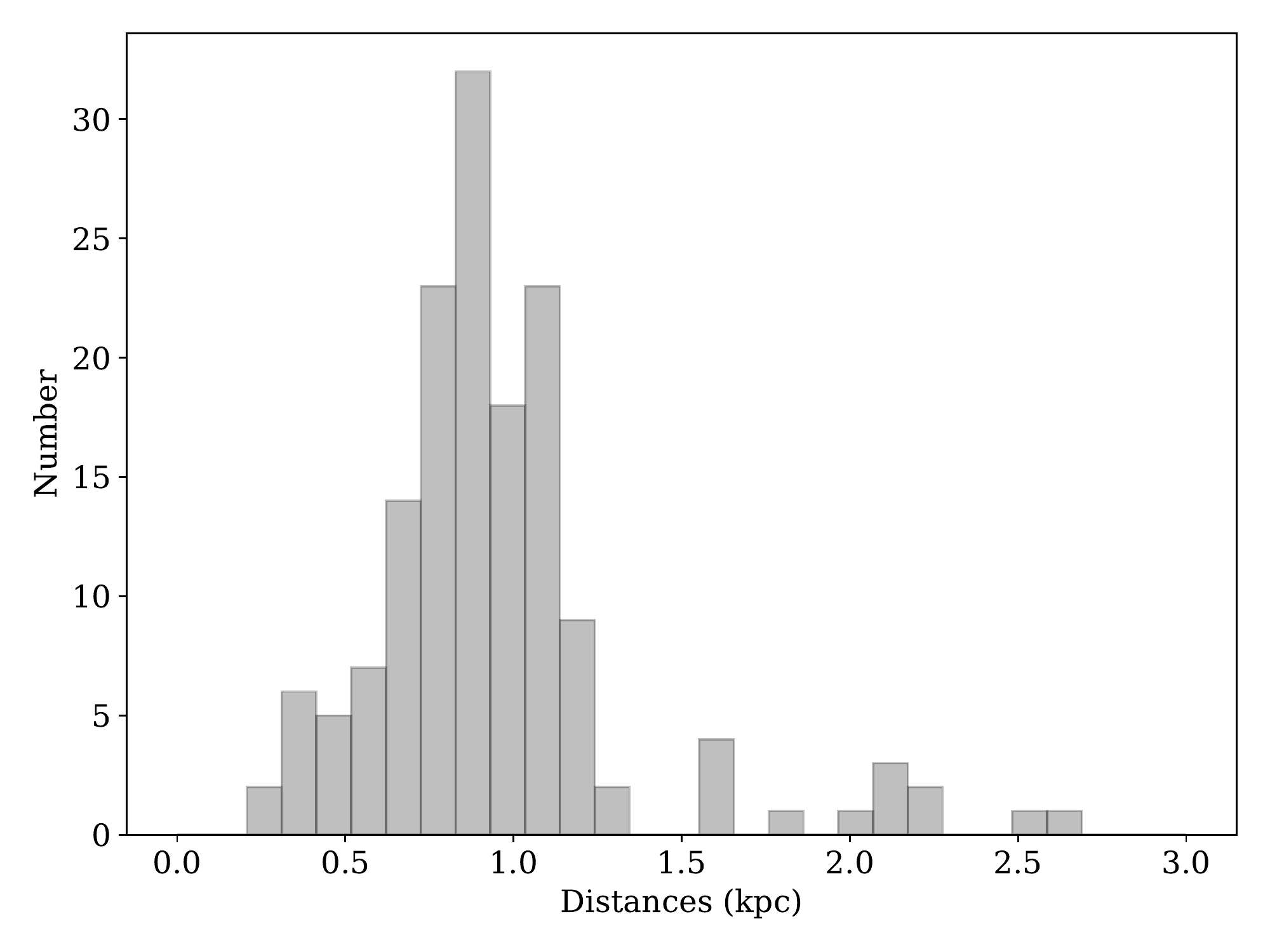}
    \caption{The distribution of the distances for the subsamples from \cite{Yan2020}. \label{fig:fdistance}}
\end{figure}

Figure \ref{fig:fdistance} presents the distribution of the distances for 154 molecular clouds. 
Their distances are measured with the extinctions and Gaia DR2 parallaxes 
through the background-eliminated extinction-parallax method in \citep{Yan2021b}. 

\clearpage
\bibliography{Sepa_13co.bib}{}
\bibliographystyle{aasjournal}



\end{document}